\documentclass[dvips,generic,preprint]{imsart}

\setattribute{journal}{name}{}

\RequirePackage[OT1]{fontenc}
\RequirePackage{amsthm,amsmath,natbib}
\RequirePackage[colorlinks,citecolor=blue,urlcolor=blue]{hyperref}
\RequirePackage{hypernat}
\RequirePackage{comment}
\RequirePackage{graphicx,subfigure,latexsym,amssymb}
\RequirePackage{float,epsfig,multirow,rotating}
\RequirePackage{upgreek,wrapfig}
\arxiv{math.PR/0000000}
\startlocaldefs

\newcommand {\ctn}{\citet} 
\newcommand {\ctp}{\citep}       

\numberwithin{equation}{section}
\theoremstyle{plain}

\newtheorem{remark}{Remark}[section]

\newcommand{\bb}{\boldsymbol{b}}

\newcommand{\boeta}{\boldsymbol{\eta}}

\newcommand{\balpha}{\boldsymbol{\alpha}}
\newcommand{\brho}{\boldsymbol{\rho}}
\newcommand{\btheta}{\boldsymbol{\theta}}

\newcommand{\bbeta}{\boldsymbol{\beta}}
\newcommand{\bdelta}{\boldsymbol{\delta}}

\newcommand{\bxi}{\boldsymbol{\xi}}

\newcommand{\bSigma}{\boldsymbol{\Sigma}}

\newcommand{\bPsi}{\boldsymbol{\Psi}}

\newcommand{\bmu}{\boldsymbol{\mu}}
\newcommand{\bzeta}{\boldsymbol{\zeta}}

\newcommand{\bOmega}{\boldsymbol{\Omega}}

\newcommand{\bB}{\boldsymbol{B}}
\newcommand{\bC}{\boldsymbol{C}}

\newcommand{\bV}{\boldsymbol{V}}

\newcommand{\bh}{\boldsymbol{h}}
\newcommand{\bH}{\boldsymbol{H}}

\newcommand{\bM}{\boldsymbol{M}}
\newcommand{\bI}{\boldsymbol{I}}

\newcommand{\bA}{\boldsymbol{A}}

\newcommand{\bQ}{\boldsymbol{Q}}

\newcommand{\bW}{\boldsymbol{W}}
\newcommand{\bu}{\boldsymbol{u}}
\newcommand{\bv}{\boldsymbol{v}}
\newcommand{\bs}{\boldsymbol{s}}
\newcommand{\bU}{\boldsymbol{U}}
\newcommand{\bx}{\boldsymbol{x}}
\newcommand{\bX}{\boldsymbol{X}}

\newcommand{\bzero}{\boldsymbol{0}}

\newcommand {\bN}{\mathcal{N}}

\newcommand {\mS}{\mathcal{S}}

\endlocaldefs

\begin{document}
\renewcommand\baselinestretch{1.}

\begin{frontmatter}
\title{A New Bayesian Test to test for the Intractability-Countering Hypothesis} 
\runtitle{New Bayesian Test of Hypothesis}

\begin{aug}
\author{
{\fnms{Dalia} \snm{Chakrabarty}\thanksref{t1,m1}\ead[label=e1]{d.chakrabarty@warwick.ac.uk}\ead[label=e2]{dc252@le.ac.uk}}
}
\thankstext{t1}{Lecturer of Statistics at Department of Mathematics,
  University of Leicester and Associate at Department of Statistics,
  University of Warwick}

\runauthor{Chakrabarty}

\affiliation{University of Warwick \and University of Leicester}

\address{\thanksmark{m1} 
Department of Mathematics\\
University of Leicester \\
Leicester LE1 7RH,
U.K.\\
\printead*{e2}\\
\and\\
Department of Statistics\\
University of Warwick\\
Coventry CV4 7AL,
U.K.\\
\printead*{e1}\\
}

\end{aug}

\begin{abstract}
{ We present a new test of hypothesis in which we seek the probability
  of the null conditioned on the data, where the null is a
  simplification undertaken to counter the intractability of the more
  complex model, that the simpler null model is nested within. With
  the more complex model rendered intractable, the null model uses a
  simplifying assumption that capacitates the learning of an unknown
  parameter vector given the data. Bayes factors are shown to be known
  only up to a ratio of unknown data-dependent constants--a problem
  that cannot be cured using prescriptions similar to those suggested
  to solve the problem caused to Bayes factor computation, by
  non-informative priors. Thus, a new test is needed in which we can
  circumvent Bayes factor computation. In this test, we undertake
  generation of data from the model in which the null hypothesis is
  true and can achieve support in the measured data for the null by
  comparing the marginalised posterior of the model parameter given
  the measured data, to that given such generated data. However, such
  a ratio of marginalised posteriors can confound interpretation of
  comparison of support in one measured data for a null, with that in
  another data set for a different null. Given an application in which
  such comparison is undertaken, we alternatively define support in a
  measured data set for a null by identifying the model parameters
  that are less consistent with the measured data than is minimally
  possible given the generated data, and realising that the higher the
  number of such parameter values, less is the support in the measured
  data for the null. Then, the probability of the null conditional on
  the data is given within an MCMC-based scheme, by marginalising the
  posterior given the measured data, over parameter values that are
  as, or more consistent with the measured data, than with the
  generated data. In the aforementioned application, we test the
  hypothesis that a galactic state space bears an isotropic
  geometry, where the (missing) data comprising measurements of some
  components of the state space vector of a sample of observed galactic
  particles, is implemented to Bayesianly learn the gravitational
  mass density of all matter in the galaxy. In lieu of an assumption
  about the state space being isotropic, the likelihood of the sought
  gravitational mass density given the data, is intractable.  For a
  real example galaxy, we find unequal values of the probability of
  the null--that the host state space is isotropic--given two
  different data sets, implying that in this galaxy, the system state
  space constitutes at least two disjoint sub-volumes that the two
  data sets respectively live in. Implementation on simulated galactic
  data is also undertaken, as is an empirical illustration on the
  well-known O-ring data, to test for the form of the thermal
  variation of the failure probability of the O-rings.
}
\end{abstract}


\end{frontmatter}

{
\section{Introduction}
\label{sec:intro}
\noindent
Model selection is a very common exercise faced by practitioners of
different disciplines, and substantial literature exists in this
field
\ctp{kassraftery,ibf_2001,chipman_2001,ghoshsamanta_2001,barabari_2004,tony,cassella2009}. In
this context, some advantages of Bayesian approaches, over
frequentist methods have been reported
\ctp{IBF_2004,robert_2001}. 
Much has been discussed in the literature to deal with the
computational challenge of Bayes factors \citep[to name a
  few]{han2000,chib2001,cassella2009}. At the same time, methods have
been advanced as possible resolutions when faced with the challenge of
improper priors on the system variables
\ctp{aitkin,IBF,tony}. Nonetheless, Bayes factor computation persists
as a challenge, especially in the context of non-parametric
and multimodal inference on a high-dimensional state space
\ctp{linkbarker_2006}.

In this paper we discuss a new test of hypothesis that is aimed at
finding support in the available data for the null that the state
space that the observed variable lives in, is endowed with a simple
symmetry, namely isotropy. In an isotropic state space, the density
at a given point depends only on the magnitude of the state space
vector to that point, and not on the inclination of this vector to a chosen
direction. This assumption about the geometry
of the state space is invoked to allow us to refer to an application, in
which the sought model parameter vector can be estimated from the
data, only under the simplistic assumption that the state space is
isotropic. In lieu of such an assumption, the likelihood of the unknown
parameters given the data is rendered intractable. Upon the estimation
of the sought parameters, given the data at hand, we want to review
how bad this assumption of isotropy of the state space is, in the
considered data. 

The application we elude to above, involves the estimation of
the density of all gravitating matter in a real galaxy NGC~3379 for
which multiple data sets are measured for two distinct types of
galactic particles \ctp{douglas_07,bergond}. The
sought model behaviour function is the gravitational mass
density function of all matter--dark as well as luminous--in this real galaxy. 
One of the burning questions in science today is the understanding of
dark matter. The quantification of the distribution of dark matter in
our Universe, at different length scales, is of major interest in
Cosmology \ctp{roberts75, rubin2001,salucci2000,deblok2003,
  hayashi2007}. At scales of individual galaxies, the relevant version
of this exercise is the estimation of the density of the gravitational
mass of luminous as well as dark matter content of these
systems. Readily available data on galactic images, can in principle
be astronomically modelled to quantify the gravitational mass density
of the luminous matter in the galaxy, \ctp{gallazi2009, bell2001};
such luminous matter is however, only a minor fraction of the total
that is responsible for the gravitational field of the galaxy since
the major fraction of the galactic gravitational mass is contributed
to by dark matter \ctp{veselina_kalinova}. Astronomical measurements
that bear signature of the gravitational effect of all (dark+luminous)
matter in a galaxy are hard to achieve in ``early-type'' galaxies, the
observed images of which is typically elliptical in
shape\footnotemark. Of some such astronomical measurements, noisy and
partially missing information on velocities of individual galactic
particles have been implemented to learn the density of all
gravitating matter in the galaxy
\ctp{cote2001,genzel,chakrabartysomak}.
\footnotetext{The intrinsic global morphology of such ``early-type''
  galaxies is approximated as a triaxial ellipsoid; in this paper we
  discuss gravitational mass density determination of this type of galaxies
  that are more frequent.}

In this application, the null states that the native space of the data variable is isotropic. This null is nested within a more complex model in
which, the data lives in a state space that is
not necessarily isotropic. However, in this application, estimation
of the model parameters is not possible under this more complex model,
given the data; in fact, even the formulation of
the likelihood of the unknown parameters given the data, is not
possible unless the null is invoked. When we refer below to the
complex model being ``intractable'', we imply the impossibility of
both formulating and computing the likelihood under this model.
Given this nature of the complex model, we find that the
posterior odds of the null model given two independent data sets is
known only upto a ratio of unknown constants, where these constants
are the uncomputable probabilities of the considered data sets. 
In form, the indeterminacy of the posterior odds appears similar to
that of the Bayes Factor when non-informative priors are used on the
model parameters--in that case, the priors are known only upto an unknown
constant, so that the the Bayes Factor is left indeterminate upto a
ratio of these unknown constants. However, unlike the indeterminacy caused by non-informative priors, the indeterminacy of the posterior odds in the considered application is entirely
data dependent, motivating us to seek a new test that bypasses computation of Bayes Factors. 
This test helps find support in a data set for a null, or can find the
ratio of supports for two nulls given two different data sets. When
the application is in the latter context, the test permits
usage of data sets of widely different sizes, and the dimensionality
of the model parameter vectors sought under the different models could
also be very different from each other. Lastly, very little prior
information may be available on the model parameter vectors in one or
both models.

This new test involves generating data from the model in which the
null is true. Though in principle, it is possible to compare the
marginalised posterior of the model parameter given measured data to
that given generated data, a ratio of these posteriors may confound
the comparison of supports in two differently sized data sets for
respective nulls, with model parameters of different
dimensionalities. Such describes the galactic application discussed
above. In such applications, support in a data for a null is given by
the probability of the null conditional on the data, which in turn is
the posterior of the model parameter marginalised over those parameter
values that are more or equally consistent with the measured data,
than is minimally achieved given the data that is generated when the
null is true.

The paper is organised as follows. Section~\ref{sec:application}
discusses the general background to the estimation of the unknown
model parameter vector and its specific formulation in the context of
the application undertaken in this work. Section~\ref{sec:null}
clarifies the formulation of the null as the assertion that the state
space that the data variable lives in, is isotropic. In this same
section we discuss the vagaries of an intractable alternative that the
null is nested within and motivate the need for a new test, which is
introduced in Section~\ref{sec:new}. Differences between this new test
and FBST are discussed in Section~\ref{sec:diffs}. An empirical
illustration of this test on the well-known O-ring data is presented
in Section~\ref{sec:oring}. The implementation of this new test in the
context of our galactic application is discussed in
Section~\ref{sec:implementation}. Such implementation is illustrated
on simulated and real data. The work with the simulated data is
presented in Section~\ref{sec:simulated} while the application to the
data of a real galaxy is included in Section~\ref{sec:RESULTS}. The
paper is concluded with a discourse on the implications of the
results, in Section~\ref{sec:discussions}.


\section{Case Study}
\label{sec:application}
\noindent
In the application that we are interested in, the state space vector
$\bW\in{\cal W}\subseteq{\mathbb R}^6,
\bW=(X_1,X_2,X_3,V_1,V_2,V_3)^T$, where $\bX=(X_1,X_2,X_3)^T$ and
$\bV=(V_1,V_2,V_3)^T$. In the application, $\bX$ is the
three-dimensional location and $\bV$ the velocity vector of a particle
in the system. The measurables include some components of $\bX$ and
some components of $\bV$--the measurable vector is
$\bU=(X_1,X_2,V_3)^T$ so that the data set is ${\bf
  D}=\{\bu_k\}_{k=1}^{N_{data}}$. Thus, $\bU\in{\cal U}\subset{\cal
  W}$. We are interested in estimating the model parameter vector
$\btheta\in{\cal S}$. In the application,
$\btheta=(\Psi_1,\ldots,\Psi_{N_{eng}},\rho_1,\rho_2,\ldots,\rho_{N_x})^T$,
where $\brho=(\rho_1,\rho_2,\ldots,\rho_{N_x})^T$ and
$\bPsi=(\Psi_1,\Psi_2,\ldots,\Psi_{N_{eng}})^T$ which respectively,
are the discretised versions of an unknown model function
$\rho(\bX)$ and the state space $pdf$. In our application, $\rho(\bX)$ is the density of gravitational mass of all (dark+luminous) matter in the galaxy, in which $\bU$ has been observed for a sample of $N_{data}$ galactic particles.

The reason for reducing our ambition from learning the full functions
$\rho(\bX)$ and the state space $pdf$, to their discretised
forms--namely $\brho$ and $\bPsi$ respectively--is the lack of
``training data'', which in this context, is the data set comprising a
set of values of the data variable $\bU$, generated at chosen values
of $\rho(\bX)$ and the state space $pdf$. However, we do not know the
physics underlying the relation between the unknown functions and
$\bU$--such is the system at hand. This results in the inability to
generate the value of $\bU$ at a chosen value of $\rho(\bX)$ and the
state space $pdf$, i.e. results in the unavailability of training data. In
this situation, we cannot take the usual approach of statistical
learning using training data, to train a model of the relationship
between the measurable and unknown functions, to thereafter predict
the unknown function by implementing the available measurements (test
data) in this model \ctp{neal1998}.

Consequently, we are left with the possibility of discretising the
support of the unknown functions and estimate the values of the
functions in each resulting grid cell, treating these values as
independent of each other without invoking a correlation
structure. Thus we can only learn the discretised forms of these
unknown functions, i.e. learn the vector $\brho$ where the $i$-th
component of $\brho$ is the value of $\rho(\bX)$ in the
$i$-th grid cell that the support of $\rho(\bX)$ is discretised into
(and likewise for the vector $\bPsi$, a component of which is the
value of the state space $pdf$ over a grid-cell, where the support
of this $pdf$ is discretised into grid-cells). 

Details of the estimation of $\brho$ and $\bPsi$ from ${\bf D}$ is
discussed in Section~{\bf S-1} of the Supplementary Material. It is to
be noted that this estimation is markedly non-trivial given that the
measurements are of parameters $X_1,X_2,V_3$ while the sought unknown
function $\rho(\bX)$ is defined over $\bX=(X_1,X_2,X_3)^T$ and the
sought unknown state space density is defined over the state space
vector $\bW=(X_1,X_2,X_3,V_1,V_2,V_3)^T$. Thus the measurables live
only in a sub-volume inside the state space, i.e. ${\cal
  U}\subset{\cal W}$. In other words, the measurables are sampled from
the density $\nu(\bU)$ of the $\bU$ vector, where $\nu(\bU)$ is
achieved by marginalising the state space density over the
non-measurables, i.e. over $X_3,V_1,V_2$. The likelihood function is
written in terms of $\nu(\bU)$ convolved with the density of the
errors in the measurables. Importantly, this likelihood is intractable
unless the state space ${\cal W}$ admits isotropy. So we assume an
isotropic state space and achieve the likelihood of the unknowns
$\brho$, $\bPsi$ given ${\bf D}$. Relevant priors are invoked and we write
the posterior of the unknowns given the data; posterior inference
is carried out using Metropolis-Hastings.

For NGC 3379, data include missing data on the three observable state
space coordinates of 164 galactic particles called planetary nebulae
(PNe)--that are the end states of certain massive stars--as reported by
\ctn{douglas_07}. In addition, there is data on 29 of another type of
galactic particles called globular clusters (GCs) that are clusters of
stars--reported by \ctn{bergond}. 


NGC 3379, or M~105, seems to have initiated its journey within the
observational domain, in neglect - though Pierre Mechain is credited
with its discovery in 1781, it did not initially make it to Messier's
catalogue. Amends were made later in 1947, when it was among four new
objects that were ``added to the accepted list of Messier's catalogue
as nos. 104, 105, 106 and 107'' (from Helen Sawyer, 1947). In spite of
this early inattention, NGC 3379 has been studied carefully in the
past few years.  \ctn{romanowskyscience} advanced the idea that NGC
3379 is one of the five ``naked galaxies'', that were tracked using
the data on the observed PNe samples in these five galaxies . Such
claims were contested by \ctn{dekel_nature}, though \ctn{douglas_07}
defend the earlier result of \ctn{romanowskyscience} by analysing the
PNe data in NGC 3379. For this galaxy, \ctn{douglas_07} also report
{\it one} value of gravitational mass at a chosen distance from the
galactic centre, obtained from using the GCs data in this galaxy
\ctp{bergond}.  This single value obtained using the GC data, is shown
to concur with the estimate based on PNe data, within error bars.
\ctn{weijmans_09} cannot infer the
distribution of the total gravitational mass distribution in this
galaxy since the halo contribution is an unknown model
parameter for them. \ctn{coccato09} and \ctn{pierce_06} report the
characterisation of this galaxy using PNe and GC data respectively.

It is to be noted that by ``training data'' in the first part of this section, we imply data that consists of pairs of design points and measurable
values generated at this design point, while in the context of Bayes
Factor literature, ``training samples'' or ``training data'' typically
imply data that mimic the available set of measurements and can
therefore be ``real'' (i.e. are samples of the available
measurements), or ``imaginary'' i.e. sampled from the posterior
predictive under the null, given the available measurements.


\section{Testing for the assumption of an isotropic state space given the data at hand}
\subsection{The null hypothesis}
\label{sec:null}
\noindent
If the state space ${\cal W}$ is isotropic, the state space density is an isotropic function of $\bX$ and $\bV$, where the state space vector is $\bW=(X_1,X_2,X_3,V_1,V_2,V_3)^T$.

\begin{remark}
\label{remark:1}
{If a real-valued function $g(\cdot,\cdot)$ of two vectors
${\bf a},{\bf b}\in{\mathbb R}^m$, is an isotropic function of ${\bf a},{\bf b}$, then
$g({\bf a, b})=g({\bf Q}{\bf a}, {\bf Q}{\bf b})$, for any orthogonal
transformation matrix ${\bf Q}\in{\mathbb R}^{(m\times m)}$ \ctp{truesdell, wang}. We recall from
the theory of scalar valued functions of two vectors, that if
$g(\cdot,\cdot)$ is an isotropic function, its set of invariants with
respect to ${\bf Q}$ is
${\Upsilon}_Q=\{{\bf a}\cdot{\bf a}, {\bf b}\cdot{\bf b}, {\bf
  a}\cdot{\bf b}\}$ where ``$\cdot$'' is the inner product of 2 vectors. Then, the isotropic function of two vectors,
$g({\bf a},{\bf b})$, admits the representation $g({\Upsilon}_Q)\equiv
g({\bf a}\cdot{\bf a}, {\bf b}\cdot{\bf b}, {\bf
  a}\cdot{\bf b})$
\ctp{truesdell, shi}. }
\end{remark}

In our application, $\bX\cdot\bV$=0 identically so that it follows
from Remark~\ref{remark:1} that if the state space density
$f(\bX,\bV)$ is an isotropic function of $\bX$ and $\bV$, then it will
depend on $\bX$ and $\bV$ via the form
$f(\bX\cdot\bX,\bV\cdot\bV)$, i.e. $f(X^2, V^2)$, since
$\bX\cdot\bX=\parallel \bX \parallel^2 =
X^2=(X_1^2+X_2^2+X_3^2)$, where $\parallel\cdot\parallel$ is the $L^2$-norm of a vector. Similarly, $\bV\cdot\bV= \parallel \bV
\parallel^2 = V^2=(V_1^2+V_2^2+V_3^2)$. Thus, in this application, any function$f(X^2,V^2)$ is an isotropic scalar-valued function of $\bX$ and $\bV$. To summarise, any function that depends on $\bX$ and $\bV$ via the $L^2$-norms of the $\bX$ and $\bV$ vectors, is an isotropic function of the 2 vectors $\bX$ and $\bV$.

In our application, it then follows that if we define a simple function of
$X$ ($:=\sqrt{X^2}={\parallel\bX\parallel}$) and
$V$ ($:={\parallel\bV\parallel}$) as: $E(X,V):=\Phi(X) +
\eta(V)$\footnotemark, the state space density that bears the form $\Psi(E)$
is an isotropic function of $\bX$ and $\bV$,
implying that state space ${\cal W}$ is isotropic. Here
$\Psi(\cdot)\geq 0$ is any function; (the constraint of
non-negativity stems from non-negativity of the state space
density). \footnotetext{In Section~\ref{sec:implementation} we will see that $\rho(\bX)$ being a known function of $\Phi(\bX)$, is embedded within the support of the state space $pdf$ under the null model, i.e. within the support of $\Psi(\cdot)$. We will then estimate the discretised version of $\rho(\bX)$ as the vector $\brho$ (as well as the discretised version $\bPsi$ of the state space density under a null model, i.e. of $\Psi(E(X,V))$).}
Thus, the null that the $i$-th data set at hand (${\bf
  D}_i$) is sampled from an isotropic state space density function
$f_i(\bX,\bV)$, is expressed as:
\begin{equation}
H_0^{(i)}: f_i(\bX,\bV) = \Psi_i(E(X,V)), \quad{\mbox{where}}\quad \Psi_i(\cdot)\geq 0,
\label{eqn:nulldefn}
\end{equation} 
where in our application, $i=1,2$. That the null model is different in the 2 cases suggests that while data ${\bf D}_1$ lives in the isotropic state space ${\cal W}_1$ under the null $H_0^{(1)}$, the data ${\bf D}_2$ does not necessarily live in the same state space but rather in a different state space ${\cal W}_2$ in general, which is isotropic under the null $H_0^{(2)}$.

We have discussed in Section~\ref{sec:application} that lack of training data causes replacement of the learning of the unknown gravitational mass density function $\rho(\bX)$ by its discretised version, namely the vector $\brho=(\rho_1,\ldots,\rho_{N_x})^T$. Similarly, the state space density function $f(\bX,\bV)=\Psi(E(X,V))$ under the assumption of isotropy, cannot be learnt, but in its place, its discretised version is learnt, namely the vector $\bPsi=(\Psi_1,\ldots,\Psi_{N_E})^T$. Then the sought model parameter vector, learnt using data ${\bf D}_i$ is $\btheta_i=(\Psi_1^{(i)},\ldots,\Psi_{N_E}^{(i)},\rho_1^{(i)},\ldots,\rho_{N_x}^{(i)})^T$, $i=1,2$.

\subsection{The alternative model is intractable}
\noindent
One would readily suggest that comparative support in data sets ${\bf
  D}_1$ and ${\bf D}_2$ for an isotropic state space (that the
respective data lives in), be given by the posterior
odds $\displaystyle{\frac{\Pr(H_0^{(1)}\vert{\bf
      D}_1)}{\Pr(H_C^{(1)}\vert{\bf D}_1)}}$ and
$\displaystyle{\frac{\Pr(H_0^{(2)}\vert{\bf
      D}_2)}{\Pr(H_C^{(2)}\vert{\bf D}_2)}}$, where the more complex model,
$H_C^{(i)}$, suggests that the $i$-th data set lives in a state space that is not necessarily isotropic; $i=1,2$. However, as we discussed above, the application is such that posterior computation under the complex model is intractable. 
In that case we could compare the posterior odds of the null and the alternative $\displaystyle{\frac{\Pr(H_0^{(1)}\vert {\bf D}_1)}{1-\Pr(H_0^{(1)}\vert {\bf D}_1)}}$ with $\displaystyle{\frac{\Pr(H_0^{(2)}\vert {\bf D}_2)}{1-\Pr(H_0^{(2)}\vert {\bf D}_2)}}$, where the alternative $H_1^{(i)}$ suggests that the $i$-data lives in an anisotropic state space, such that $\Pr(H_1^{(i)}\vert {\bf D}_i) = 1 - \Pr(H_0^{(i)}\vert {\bf D}_i)$.
Now, from Bayes rule, we can express the posterior of $H_0^{(i)}$ given the $i$-th data set, as proportional to the likelihood of this null given data ${\bf D}_i$ and the prior on this null, so that
\begin{equation}
\displaystyle{\frac{\Pr(H_0\vert {\bf D}_i)}{1-\Pr(H_0\vert {\bf D}_i)}} = \displaystyle{\frac{\alpha_i\Pr({\bf D}_i\vert H_0)\Pr(H_0)}{1-\alpha_i\Pr({\bf D}_i\vert H_0)\Pr(H_0)}}
\label{eqn:postodds1}
\end{equation}
where $\alpha_i$ is defined as the reciprocal of $\Pr({\bf D}_i)$, i.e.
\begin{equation}
\alpha_i^{-1}:= \Pr({\bf D}_i)=\Pr({\bf D}_i\vert H_0^{(i)})\Pr(H_0^{(i)}) + 
\displaystyle{\sum\limits_{j}\Pr({\bf D}_i\vert H_{1j}^{(i)})\Pr(H_{1j}^{(i)})},
\label{eqn:alphadefn}
\end{equation}
showing the probability of the data ${\bf D}_i$ at hand as conditional
upon an isotropic model for the state space (1st term on RHS of
Equation~\ref{eqn:alphadefn}), and upon all possible disjoint
anisotropic models $H_{1i}$ for the state space (2nd term on RHS). Then $\alpha_i$ cannot be computed,
since this 2nd term on the RHS of Equation~\ref{eqn:alphadefn} cannot
be computed. This is because, likelihood under the anisotropic model given the data
is not computable due to the intractability of the anisotropic
model. This then implies that the posterior odds expressed in
Equation~\ref{eqn:postodds1} is not known.

In fact, we find that if we express the posterior odds of null $H_0^{(1)}$ given data ${\bf D}_1$ to $H_0^{(2)}$ given ${\bf D}_2$, such an odds ratio is known only upto the ratio of the unknown constants $\displaystyle{\frac{\alpha_1}{\alpha_2}}$, as in the following. 
\begin{equation}
\displaystyle{\frac{\Pr(H_0^{(1)}\vert{\bf
      D}_1)}{\Pr(H_0^{(2)}\vert{\bf D}_2)}} = 
\displaystyle{\frac{\alpha_1}{\alpha_2}} \times \displaystyle{\frac{\Pr({\bf D}_1\vert{H_0^{(1)}})}{\Pr({\bf D}_2\vert{H_0^{(2)}})}}\times \displaystyle{\frac{\Pr(H_0^{(1)})}{\Pr(H_0^{(2)})}},
\label{eqn:postodds2}
\end{equation}
where $\alpha_i$ is unknown, $i=1,2$, so that the indeterminacy in the
posterior odds in Equation~\ref{eqn:postodds2} is due to the unknown
ratio $\alpha_1/\alpha_2$. (We stress that the 2nd factor on the RHS
of Equation~\ref{eqn:postodds2} is not the Bayes Factor since it is
the ratio of marginals of two different data sets, given the
respective null). Yet, the form of this indeterminacy is reminiscent
of the form of the indeterminacy in Bayes Factors (BFs) when one uses
non-informative priors on the unknown model parameter vector $\btheta$
such that these priors are known only upto an unknown constant--we can
then compute BFs in principle, with posterior Bayes factors
\ctp{aitkin}, intrinsic Bayes factors \ctp{IBF,ibf_1996a} or with
fractional Bayes factors \ctp{tony}.  We clarify this similarity in
form between the two indeterminacies in the following section.

\subsection{Indeterminacy of Bayes Factors given non-informative priors and irrelevance of prescribed cures to our posterior odds}
\label{sec:indeterminacy}
\noindent
The posterior odds of the two null models given the respective data
sets, is expressed
in Equation~\ref{eqn:postodds2}. Now, we can set the prior odds for
the nulls $H_0^{(1)}$ and $H_0^{(1)}$ to be unity and rewrite the
posterior odds $\displaystyle{\frac{\Pr(H_0^{(1)}\vert{\bf
      D}_1)}{\Pr(H_0^{(2)}\vert{\bf D}_2)}}$ by expanding the marginal
likelihood given data set ${\bf D}_i$ in terms of the likelihood
$f_i({\bf D}_i\vert\btheta_i)$ of the unknown model parameter
$\btheta_i$ given this data, and the prior $\pi_0(\btheta_i)$ of
$\btheta_i$. Here we realise that the model parameter vector sought
under the model $H_0^{(1)}$ is not equal to that sought under the
model $H_0^{(2)}$; hence these parameters are distinguished in the
notation as $\btheta_1$ and $\btheta_2$. Likewise, the notation
acknowledges for difference between the likelihood function of the
unknown parameter given one data set in one case, and the other given the other dataset in the other case. Thus under prior odds of unity, i.e. for 
$\Pr(H_0^{(1)})=\Pr(H_0^{(2)})$,
\begin{equation}
\displaystyle{\frac{\Pr(H_0^{(1)}\vert{\bf
      D}_1)}{\Pr(H_0^{(2)}\vert{\bf D}_2)}} = 
\displaystyle{                                                                 
\frac{\alpha_1\Pr({\bf D}_1\vert H_0^{(1)})\Pr(H_0^{(1)})}               
     {\alpha_2\Pr({\bf D}_2\vert H_0^{(2)})\Pr(H_0^{(2)})}                
                         } =                                                  
  \displaystyle{\frac{\alpha_1\int f_1({\bf D}_1\vert \btheta_1)\pi_0(\btheta_1)d\btheta_1}{\alpha_2\int f_2({\bf D}_2\vert \btheta_2)\pi_0(\btheta_2)d\btheta_2}}
\label{eqn:postodds3}
\end{equation}
Then Equation~\ref{eqn:postodds3} indicates that if the prior on $\btheta_i$ is non-informative, so that it is known only upto an unknown constant $c_i$, then the indeterminacy in the posterior odds is compounded by the factor $\displaystyle{\frac{c_1}{c_2}}$ in addition to the existing indeterminacy due to the unknown ratio $\displaystyle{\frac{\alpha_1}{\alpha_2}}$. 

The problem about BFs being known upto the ratio of the unknown
constants $c_1/c_2$ that stems from the usage of non-informative
priors on the model parameters, has been dealt with in the literature
\ctp{IBF_2004}. In this situation, the BF is the ratio given the
models 1 and 2 and is arbitrary in its scale; here this ``arbitrary
BF'' is ${\cal B}_{12}^{(A)}:=\displaystyle{\frac{c_1\int f_1({\bf
      D}\vert \btheta_1)\pi_0(\btheta_1)d\btheta_1}{c_2\int f_2({\bf
      D}\vert \btheta_2)\pi_0(\btheta_2)d\btheta_2}}$. (We note that
the BF having been defined at a given data set, is not quite the ratio
of the marginal likelihoods given the 2 different data sets that we
consider in Equation~\ref{eqn:postodds3}). The suggestion that is
offered in the literature is that ${\cal B}_{12}^{(A)}$, needs to be
replaced by the fully computable BF ${\cal B}_{12}$ where ${\cal
  B}_{12}$ is defined as:
${\cal B}_{12}={\cal B}_{12}^{(A)}\left\langle {\cal B}_{21}^{(A)}({\bf D}^{(\ell)})\right\rangle$,
where ${\cal B}_{12}^{(A)}$ is computed using the available data ${\bf
  D}$ while $\left\langle{\cal B}_{21}^{(A)}({\bf
  D}^{(\ell)})\right\rangle$ is the average computed using the new
data set ${\bf D}^{\ell}$, with the averaging performed over all such
``new''--or training data. Indeed, the indeterminacy in the BF caused
by the ratio $c_1/c_2$ is eliminated in this prescription.
As mentioned in Section~\ref{sec:application}, training data could typically
imply data that mimic the available set of measurements and can
therefore be ``real'' (i.e. ${\bf D}^{(\ell)}$ is one partition of the
available measurements), or ``imaginary'' i.e. sampled from the
posterior predictive under the null, given the available measurements
\ctp{IBF}. The posterior of the model parameter $\btheta_i$ given an
``imaginary'' ${\bf D}^{(\ell)}$, averaged over all ${\bf
  D}^{(\ell)}$, is then referred to as the ``expected-posterior
prior'' of $\btheta_i$ under the null $H_0^{(i)}$, and used in place
of the prior on $\btheta_i$, according to $\displaystyle{\int
  \pi(\btheta_i \vert {\bf D}_i^{(\ell)}) m_i({\bf D}_i^{(\ell)})
  d{\bf D}_i^{(\ell)}}$--see \ctn{fouskakis}. Here $m_i({\bf
  D}_i^{(\ell)}) = \displaystyle{\int{f_i({\bf D}_i^{(\ell)}\vert
    {\btheta}_i)\pi_0(\btheta_i)d{\btheta_i} }}$.

Irrespective of the nature of the training data, the prescription that
helps cure the indeterminacy caused by the usage of non-informative
priors on $\btheta_i$, i.e. the data-independent unknowns
$c_i$. However it is irrelevant to curing the indeterminacy in the
posterior odds of Equation~\ref{eqn:postodds3} that is caused by the
uncomputable data-dependent ratio $\displaystyle{\frac{\Pr({\bf
      D}_2)}{\Pr({\bf
      D}_1)}}\equiv\displaystyle{\frac{\alpha_1}{\alpha_2}}$ , where
the uncomputable nature of this probability owes to the intractable
nature of the complex model that the $i$-th null is nested within, $i=1,2$. It
is then clear that multiplying the ratio of the marginal likelihoods
of the data under the respective null, by its reciprocal
computed at new data sets ${\bf D}_1^{(\ell)}$ and ${\bf
  D}_2^{(\ell)}$, will only introduce a new ratio of unknowns
$\displaystyle{\frac{\Pr({\bf D}_1^{(\ell)})}{\Pr({\bf
      D}_2^{(\ell)})}}$ to compound the problem.
\subsection{Tractable alternative--numerical difficulties in high dimensions}
\noindent
The new test that we discuss herein, is relevant even when the complex
model that the simpler null is nested within is tractable--unlike in
the galactic application we consider here--though it is challenging in a
high-dimensional non-parametric situation, to achieve intrinsic priors
with imaginary training data sets \ctp{IBF}, or where real training
data are unachievable given that the available measurements are
under-abundant to begin with. Implementation of imaginary training
data sets may be hard when $\btheta$ is high dimensional; the
computational intricacy involved in averaging over all possible
imaginary samples would increase with increase in dimensionality of
$\btheta$. We would need to generate a large sample of training data
sets, and for each these training data sets, we would need to learn
the high-dimensional $\btheta_1$ under the null $H_0^{(1)}$ and
$\btheta_2$ under $H_0^{(2)}$.  This suggests running twice as many,
long MCMC chains to convergence, as there are training data sets that
are averaged over. This is required to be a large number, if we want
to explore the expected non-linearity in the joint posterior
probability of the large number of components of the high-dimensional
$\btheta_i$. Given such a computationally intensive method, we seek a
new method that is numerically less cost intensive.

\section{The new test}
\label{sec:new}
\noindent
In the new test we express the support in the measured data ${\bf
  D}_i$ for the null $H_0^{(i)}$, without invoking the ratio of
posterior under the null and the more complex
model--to be precise, we compute the probability of the null
hypothesis, conditional on the measured data, by marginalising the
posterior of the model parameter $\btheta_i$ given ${\bf D}_i$, over
all those $\btheta_i$ that are at least as consistent with the data,
as is minimally possible when the null is true. The posterior when
the null is true, is computed as the posterior of $\btheta_i$ given
data ${\bf D}^{/}_i$, where ${\bf D}^{/}_i$ is such that
$\Pr(H_0^{(i)}\vert {\bf D}^{/}_i) = 1$.  In other words, ${\bf
  D}^{/}_i$ is the data that is generated from the model in which the
null $H_0^{(i)}$ is true, and is referred to as the
``generated data''--to be distinguished from the measured data ${\bf
  D}_i$, i.e. generated data ${\bf D}^{/}_i$ is different from
available measured data ${\bf D}_i$, in general. Then the posterior
probability density of $\btheta_i$ given the generated data ${\bf
  D}^{/}_i$ is its posterior if the null were true. Hereafter, we
refer to this model that the null is true in, as the ``benchmark
model'' and denote it by the notation ${\cal M}_i$. For example, in
the galactic application considered in this paper, the benchmark model
is one in which the state space $pdf$ is an isotropic function of the
location and velocity vectors.

When the posterior probability of the $i$-th model parameter
$\btheta_i$ can be computed given the $i$-th measured data, as well as given
the $i$-th generated data--even if the same non-informative prior is invoked
in each posterior computation--it may be possible to define the
support in this measured data for the $i$-th null, by comparing the
marginalised posterior of $\btheta_i$ given the measured data ${\bf D}_i$, to the marginalised posterior of $\btheta_i$ when the $i$-th null is true, i.e. by comparing
$\displaystyle{\int\limits_{\btheta_i}\pi(\btheta_i\vert{\bf D}_i) d\btheta_i}$, to
$\displaystyle{\int\limits_{\btheta_i}\pi(\btheta_i\vert{\bf D}^/_i) d\btheta_i}$. In other words, the support in this
measured data for this null {\it could} in principle be given by the
odds ratio
\begin{equation}
\displaystyle{\Omega_i}=
\displaystyle{\frac
    {\int\limits_{\btheta_i} \pi(\btheta_i\vert{\bf D}_i) d\btheta_i}
    {\int\limits_{\btheta_i} \pi(\btheta_i\vert{\bf D}^/_i) d\btheta_i}
},
\label{eqn:oddsratio}
    \end{equation}
(where $i=1,2$ in our galactic application).
In that case, an odds ratio $\Omega_i\geq 1$ would imply that the support in the measured data for the null is high, with higher support for bigger values of the ratio. Similarly, $\Omega_i < 1$ would indicate lower support. However, such a definition of the support for the null in the data,
could confound the interpretation of the comparison of support in
measured data ${\bf D}_1$ for null $H_0^{(1)}$, with support in
another measured data set ${\bf D}_2$ for null $H_0^{(2)}$, where the two data sets are differently sized and the model parameters are of different dimensionalities--a
comparative exercise of this nature is the prime target in this work,
insofar as the galactic application is concerned. Such a comparison is
easier to interpret if the defined support in a data for a null is
bounded from both ends. To achieve the same, we opt to define the
support in the measured data for the null, as the probability of the
null conditional on the data, i.e. as $\Pr(H_0^{(i)}\vert{\bf
  D}_i)$. In this definition then, there can be zero support in the
data for the null
while the maximal support is 1, s.t. there is no distinction made in
this definition, between models that offer odds ratio (defined in
Equation~\ref{eqn:oddsratio}) in excess of 1. Then the support in
${\bf D}_1$ for $H_0^{(1)}$ is easily compared to that in ${\bf D}_2$
for $H_0^{(2)}$, as $\Pr(H_0^{(1)}\vert {\bf D}_1)/\Pr(H_0^{(2)}\vert
{\bf D}_2)$. However, when the application does not involve comparison of
supports in two different data sets, for respective nulls, the odds
ratio $\Omega$ of Equation~\ref{eqn:oddsratio} is indeed applicable (as in the
example application on the O-ring data, presented in
Section~\ref{sec:oring}). The pursuit of the definition of support as
the probability of the null conditional on the data--as distinguished
from the odds ratio--may appear to resemble the Fully Bayesian
Significance Test or FBST \cite{fbst2008}. FBST tests the sharp null
hypothesis that the relevant model parameter $\beta$, has a value
$\beta_0$, i.e. $H_0: \beta=\beta_0$. We discuss FBST in detail in
Section~{\bf S-2} of the attached Supplementary Material. However,
this new test differs from FBST in both scope (allows for
implementation to non-sharp nulls, in high-dimensional, non-parametric
contexts), as well as in structure (by invoking posterior computation
given the generated data, unlike by identifying the posterior computed
at the null-abiding value $\beta_0$ of the model parameter, as in
FBST). These differences are clarified in Section~\ref{sec:diffs}.  In
our definition of support as the probability of the null given the
data, we partition the native space of model parameter $\btheta_i$
into the space ${\cal T}_{{\cal M}_i}({\bf D}_i)$ that harbours
parameters that are more or equally consistent with the measured data
than is minimally possible when the null is true, and compute
$\Pr(\btheta_i\in {\cal T}_{{\cal M}_i}({\bf D}_i))$. We discuss this
construct in the following paragraphs.

Let $\btheta_i\in{\cal S}_i\subseteq{\mathbb R}^d$. We begin by partitioning 
${\cal S}_i$ into the data-dependent, disjoint and exhaustive sub-spaces ${\cal T}_{{\cal
    M}_i}({\bf D}_i)$ and $\overline{{\cal T}_{{\cal M}_i}({\bf
    D}_i)}$, for a given benchmark model ${\cal M}_i$, such that
${\cal S}_i={\cal T}_{{\cal M}_i}({\bf D}_i) \cup {\overline{{\cal
      T}_{{\cal M}_i}({\bf D}_i)}}$ where for $\btheta_i\in
\overline{{\cal T}_{{\cal M}_i}({\bf D}_i)}$, $\pi(\btheta_i\vert {\bf
  D}_i)$, is less than the minimum value of $\pi(\btheta_i\vert{\bf
  D}^{/}_i)$, i.e. the minimum value of the posterior if the null
$H_0^{(i)}$ were true. Again, for $\btheta_i\in {\cal T}_{{\cal
    M}_i}({\bf D}_i)$, $\pi(\btheta_i\vert {\bf D}_i)$, is equal to, or
in excess of the minimum value of $\pi(\btheta_i\vert{\bf D}^{/}_i)$.
In other words, ${\cal T}_{{\cal M}_i}({\bf D}_i)$ contains all
$\btheta$ that are at least as consistent with the measured data ${\bf
  D}_i$ as is minimally possible if the null were true and
$\overline{{\cal T}_{{\cal M}_i}({\bf D}_i)}$ contains all $\btheta$
that are less consistent with the measured data ${\bf D}_i$ than is
minimally possible if the null were true. The larger the proportion
of $\btheta$ that live in $\overline{{\cal T}_{{\cal M}_i}({\bf
    D}_i)}$, the smaller is the support in data ${\bf D}_i$ towards
the null. Then we can express the conditional probability
$\Pr(H_0^{(i)}\vert {\bf D}_i)$, as $1-\Pr(\btheta\in\overline{{\cal
    T}_{{\cal M}_i}({\bf D}_i)}$, which in turn is the probability that
$\btheta$ lives inside ${\cal T}_{{\cal M}_i}({\bf D}_i)$:
\begin{eqnarray}
\Pr(H_0^{(i)}\vert {\bf D}_i) &=& \Pr(\btheta_i\in {\cal T}_{{\cal M}_i}({\bf D}_i))\quad{\mbox{where}}
\label{eqn:support}\\
\Pr(\btheta_i\in {\cal T}_{{\cal M}_i}({\bf D}_i)) &=&
\displaystyle{\int\limits_{{\cal T}_{{\cal M}_i}({\bf D}_i)} \pi(\btheta_i\vert{\bf D}_i)d\btheta_i} \quad{\mbox{with}}
\label{eqn:intermsofpost}\\
{\cal T}_{{\cal M}_i}({\bf D}_i) &=& \displaystyle{
\left\{\btheta: 
\frac{\pi^{(min)}(\btheta\vert {\bf D}^{/}_i)}{r(\btheta)} \leq 
\frac{{\pi(\btheta|{\bf D}_i)}}{r(\btheta)} \right\}},
\label{eqn:T}
\end{eqnarray}

where $\pi^{(min)}(\btheta_i\vert {\bf D}^{/}_i)$ is the minimum value
of the posterior probability density of the unknown model parameter
vector $\btheta_i$ if the null were true, i.e. in the benchmark model
${\cal M}_i$. Actually, to ensure invariance to a bijective and continuously
differentiable transformation $\Xi(\cdot)$ of $\btheta_i$, in
Equation~\ref{eqn:T}, we define ${\cal T}_{{\cal M}_i}({\bf D}_i)$ as
the set of all $\theta_i$'s, the normalised posterior density of which
given data ${\bf D}_i$ is greater than or equal to the normalised
posterior under the benchmark model, with the normalisation given by a
reference density $r(\btheta_i)$, $r:{\cal S}_i\longrightarrow
{\mathbb R}$. We choose to work with a reference density
$r(\btheta_i)$, that is uniform in $\btheta_i$, $i=1,2$. Then using
this normalisation, $\Pr(H_0^{(i)}\vert {\bf D}_i)$ is rendered
invariant to re-parametrisation of $\theta_i$ brought about by the
transformation $\omega=\Xi(\btheta_i)$, \ctp{madurgajisp}; the
authors presented this suggestion in the context of FBST \ctp{fbst2008}. 


Thus Equation~\ref{eqn:support}, Equation~\ref{eqn:intermsofpost} and
Equation~\ref{eqn:T} tell us that in this new test, the definition of
the sub-space ${\cal T}_{{\cal M}_i}({\bf D}_i)$ follows from the
identification of the minimal posterior probability density of
$\btheta_i$ given generated data ${\bf D}^{/}_i$, achieved if the null
were true, i.e. achieved in the benchmark model ${\cal M}_i$. Once the
sub-space ${\cal T}_{{\cal M}_i}({\bf D}_i)$ is identified for a
chosen ${\cal M}_i$, support in ${\bf D}_i$ for null ${H}_0^{(i)}$ is
quantified by integrating the posterior density over all the
$\btheta_i$ that live inside ${\cal T}_{{\cal M}_i}({\bf D}_i)$. Thus,
unlike in Bayes Factors--the computation of which involves integrating
over the whole of the parameter space ${\cal S}_i$--this test involves
integrating over an identified subspace, ${\cal T}_{{\cal M}_i}({\bf
  D}_i)$ of ${\cal S}_i$.  

In practice, $\Pr(\btheta_i\in {\cal
  T}_{{\cal M}_i}({\bf D}_i))$ is approximated as the proportion of
samples of $\btheta_i$ generated in the MCMC chain run with measured
data ${\bf D}$, that exceed the minimal posterior attained in the MCMC
chain run with generated data ${\bf D}^/$. It is this proportion of
parameter values that reside in the subspace ${\cal T}_{{\cal
    M}_i}({\bf D}_i)$, and so, this is the proportion of values of
$\btheta_i$ that are at least as consistent with data ${\bf D}_i$,
than is minimally possible if the null were true. The conditional
probability of the null given the measured data, is then the computed
$\Pr(\btheta_i\in {\cal T}_{{\cal M}_i}({\bf D}_i))$.

Once we know how to compute
the probability of a null conditional on the measured data, 
we can compute probability of nulls $H_0^{(1)}$ and $H_0^{(2)}$
respectively, given data ${\bf D}_1$ and ${\bf D}_2$. To do this we
would need to generate data ${\bf D}^{/}_1$ and ${\bf D}^{/}_2$ from
benchmark models ${\cal M}_1$ and ${\cal M}_2$ respectively, where,
the benchmark model ${\cal M}_1$ is defined such that in it null
$H_0^{(1)}$ is true, while model ${\cal M}_2$ can be defined so that
null $H_0^{(2)}$ is true. Then we can finally compare
$\Pr(H_0^{(1)}\vert {\bf D}_1)$ with $\Pr(H_0^{(2)}\vert {\bf D}_2)$.
In fact in our galactic application--as we shall see below--${\bf
  D}_i^{/}$ is the data generated by sampling from the isotropic state
space $pdf$ that is learnt using the measured data ${\bf D}_i$;
$i=1,2$. The benchmark model ${\cal M}_i$ is then the model in which
the $i$-th state space $pdf$ is isotropic, i.e. null $H_0^{(i)}$ is
true; $i=1,2$. As mentioned at the end of Section~\ref{sec:null}, in
this application, we learn the unknown model parameter
vector $\btheta_i:=(\Psi_1^{(i)},\ldots,\Psi_{N_E}^{(i)},\rho_1^{(i)},\ldots,\rho_{N_x}^{(i)})^T$,
using the data ${\bf D}_i$, $i=1,2$. Then the support in the data
${\bf D}_i$ for the null that state space ${\cal W}_i$ is isotropic,
is given by $\Pr(H_0^{(i)}\vert {\bf D}_i) = \Pr(\brho_i,\bPsi_i\in
{\cal T}_{{\cal M}_1}({\bf D}_i))$, $i=1,2$. In
Section~\ref{sec:implementation} we discuss the implementation of this
new test to find such support in
\begin{itemize}
\item 2 data sets of disparate sizes,
\item when it is not possible to learn $\btheta_i$ under the consideration that the $i$-th data lives in an anisotropic state space for $i=1,2$ (since such an alternative model is intractable),
\item when $\btheta_1$ and $\btheta_2$ have different dimensionalities, and 
\item the error distributions of the measurables $X_1,X_2,V_3$ in data ${\bf D}_1$ and ${\bf D}_2$ are not the same.
\end{itemize}
It is to be noted that marginalisation is undertaken in this new test, as in Bayes factor computation, but unlike with BFs, the marginalisation is not over the full parameter space. Instead the marginalisation is over that sub-space of the parameter space that harbours those model parameter values that are more or equally compatible with the available data, than with the generated data, i.e. than when the null is true. In seeking such a sub-space, there is a motivational similarity in this procedure with FBST, though there are structural differences between FBST and the computation of support in our test. These are discussed in the next subsection.

Before proceeding to discuss those differences, we note that definition for support in the data for a null as per Equation~\ref{eqn:support}, is not an approximation for Bayes factors in any sense. One worry about this implementation--alluded to early in this section--is that there is no distinction made between models that enjoy support of 1 in the data given the null. On the contrary, the odds ratio computed as marginalisation over the full parameter space given the measured and generated data (Equation~\ref{eqn:oddsratio}), when applicable, is capable of distinguishing between all models that are differently compatible with the data. In applications that cannot be addressed by Bayes factors, or by the odds ratio computation, computation of support as per Equation~\ref{eqn:support} is a good way out, but there may remain worries about its asymptotic consistency. 

\subsection{Differences with FBST}
\label{sec:diffs}
\noindent
This new test differs from FBST as far as its remit as well as its
structure is concerned. 

In FBST, one seeks the maximum value of the posterior of
the model parameter $\beta$ given the available data ${\bf D}$, computed at
the value $\beta_0$ of the  model parameter, since the (sharp) null states that $\beta=\beta_0$. Then the probability that the posterior of the model parameter given ${\bf D}$ exceeds or equals this identified maximal value, is used to 
compute the support in the null given the data. However, in our new test,
the instrument of use is the ``generated data'', i.e. the data that is
generated from the model in which the null is true. With the generated
data in hand, there is no need to evaluate the posterior of the model
parameter $\btheta$ given the measured data, at chosen values of
$\btheta$. Rather, it is the posterior of $\btheta$ given ${\bf D}$, that is effectively compared to the posterior of $\btheta$ given the generated data. Consequently, even if the null is not sharp, but states
that the data is chosen from a density with a certain symmetry/form, we
can still test for such a null in ${\bf D}$. An example of this is the very galactic application that we address
in this paper. We recall from Section~\ref{sec:null} that in this
application, the null states that the host space of the state space
vector $\bW=(X_1,X_2,X_3,V_1,V_2,V_3)^T$ is isotropic. This is
inherently a non-sharp hypothesis--we express this null in a form that
may appear sharp, but only speciously so, by stating that the state
space density $f(\bX,\bV)$ is an isotropic function of $\bX$ and $\bV$
under the null, i.e. $H_0:f(\bX,\bV)=\Psi(E(X,V))$, where
$\Psi(\cdot)$ can be any function, as long as $\Psi(\cdot) \geq 0$
(see Equation~\ref{eqn:nulldefn}). Thus, in contrast to the sharp
hypothesis that states that the model parameter $\beta$
equals a known value $\beta_0$, our null states that the state space
density enjoys a prescribed symmetry, namely isotropy, and not a particular value,
since the value of the function $\Psi(E)$ is not fixed. The benchmark
model in which this null is true, is then one in which the state space
density is assumed to be an isotropic function of $\bX$ and $\bV$,
without any further specification. In fact, we undertake an empirical
illustration of our test in the following subsection, to demonstrate
that the new test can compute support in a measured data set for a
diffused null that states that the data is described by a model function that is an approximation for a known descriptor of the data, where the quality of this approximation is given. Such applications are outside the remit of FBST in its current form. Thus, one prime difference between the new test and FBST is that this test finds
support in the measured data for a hypothesis that is not necessarily
sharp, while FBST is limited to hypothesis of the type
$H_0:\theta=\theta_0$, i.e. sharp hypotheses.

In this test we can even compute support in the measured data for the
null as the ratio of the marginalised posteriors computed given the
measured and generated data--except, such a construct is difficult to
interpret when we seek to compare support in one data for a given
null, to support in another data for another null. Indeed, in
applications that do not involve such a comparison, using our test, we
can compute support in the data for a null either as $\Pr(\btheta\in
{\cal T}_{{\cal M}}({\bf D}))$, or as the odds ratio
$\Omega$ defined in Equation~\ref{eqn:oddsratio}. This is
undertaken in our empirical illustration discussed in the following
section. However, in the galactic application, we do undertake a
comparison of supports for different nulls in respective data sets,
and therefore, support in the $i$-th data for the $i$-th null is
computed only as $\Pr(\btheta_i\in {\cal T}_{{\cal M}_i}({\bf D}_i))$.  

In such applications, we identify the minimal posterior attained if
the null were true, i.e. given the generated data, and compute the
probability that this minimal value is equalled or exceeded by the
posterior of $\btheta$ given ${\bf D}$. In this pursuit, there is a motivational similarity between our test and FBST. However, unlike in FBST, computation of
this probability is performed by counting the fraction of samples of
$\btheta$ generated in the MCMC chain run with ${\bf D}$, for which
the posterior exceeds the minimal posterior attained in the 
MCMC chain run with the generated data--thus avoiding an
explicit $\arg(\max(\cdot))$ of the posterior given the generated
data. Importantly, avoiding such optimisation then helps us to extend the
applicability of this test to contexts in which $\btheta$ is high-dimensional (as borne by the galactic application). In contrast, undertaking such optimisation under the null in FBST, will get more difficult with increasing dimensionality of the model parameter, thus limiting the applicability of FBST to low-dimensional contexts.

Implementation in this new test also helps enhance its applicability
over FBST, to non-parametric situations, i.e. when the posterior
probability of $\btheta$ given data (measured and/or generated) is not
closed-form, as well as when the model in which the null is true, is
not parametric, as demonstrated by our galactic application--such a
non-parametric application is outside the scope of FBST in its current
form.

\subsection{Illustration using standard data for a diffused null}
\label{sec:oring}
\noindent
We illustrate the new test using a simple and standard data set, before moving on to implementing it on galactic data. For the purposes of this illustration, we invoke the well-known (though potently morbid) data on the failure of O-rings with temperature, \ctp{oring, casella_bk}. The ``O-rings'' are the rubber rings that were used to seal the joints in a part of the Challenger space shuttle, that exploded on the 28th of January, 1986, within a little more than the first minute of its flight. The explosion was attributed to the failure of an O-ring in this part, where O-ring failure is now known to be induced at low temperatures, such as the very low temperature of 31$^\circ$ F at the time of the Challenger launch. 

The data that we use here is the same given on page 15 of the book by \ctn{casella_bk}. This data set includes the temperature $T$ (in $^\circ$ F) at the time of the flight and the corresponding O-ring failure or success--given as 1 or 0, respectively--in 23 shuttle flights. Logistic regression is a natural choice to model the effect of the predictor variable $T$ on this binary predictor $Y$ of O-ring failure. \ctn{casella_bk} treat $Y\sim {\textrm{Bernoulli}}(p(T))$, where the rate $p(T)$ of this Bernoulli distribution is temperature dependent, with $\displaystyle{\log\left(\frac{p(T)}{1-p(T)}\right)} = \alpha + \beta T$, so that $p(T) =\displaystyle{\frac{e^{\alpha + \beta T}}{1 - e^{\alpha+\beta T}}}$, where $\alpha, \beta$ are the parameters of this logistic regression model, to be learnt given the O-ring data. Then the likelihood function is
\begin{equation}
\ell(\alpha,\beta)=\displaystyle{
{{\prod\limits_{i=1}^{23}}} 
\left(p_i\right)^{y_i} \left(1 - p_i\right)^{1-y_i}
},
\label{eqn:logis}
\end{equation}
where in the O-ring data, at the temperature $T=t_i$ in the $i$-th row, $Y=y_i$ with probability of failure given by $p_i$; $i=1,\ldots,23$. (Temperature $T\in\tau\subset{\mathbb R}$; by writing $T=t_i$, we imply a temperature in the $\epsilon$-neighbourhood of $t_i$, in the limit of $\epsilon$ approaching zero). With this likelihood, and chosen priors on $\alpha$ and $\beta$, \ctn{casella_bk} express the posterior probability density of these parameters given the O-ring data, from which they perform posterior sampling using Metropolis-Hastings (independent sampler), to learn $\alpha$ and $\beta$. 
At the modes of the marginal posterior probability of $\alpha$ and $\beta$, (at approximately 15.25 and -0.24 respectively), the $p_i$ values computed in this logistic model for $i=1,\ldots,23$, are plotted in filled black circles in Figure~\ref{fig:ptemp}, and the learnt function $p(T)$ in this model is depicted by the solid black line that connects these points in this figure. We refer to this model of $p(T)$ as $p_{mode}(T)$--to signify that this model is achieved using the modal values of $\alpha$ and $\beta$ learnt by \ctn{casella_bk}, given the O-ring data ${\bf D}=\{y_1, \ldots, y_{23}\}$.

\begin{figure*}[!t]
     \begin{center}
  {
\includegraphics[height=8.5cm]{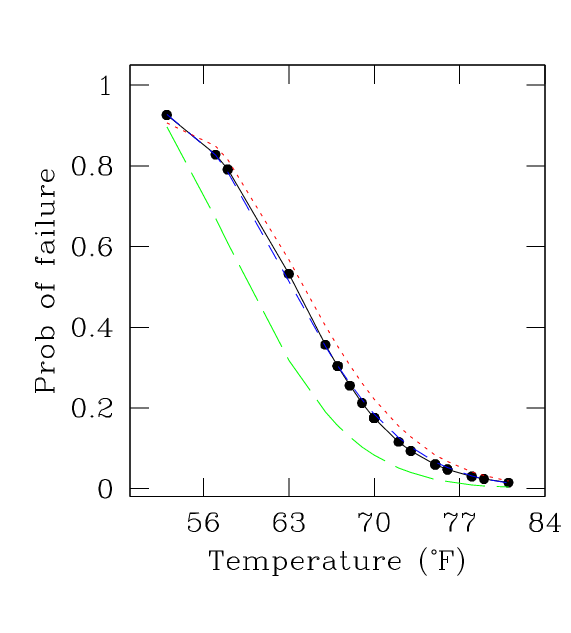} 
}
\end{center}
\vspace{-1cm}
\caption{The solid black line shows failure probability variation $p_{mode}(T)$ with temperature $T$, as learnt using the modal values of the parameters of the logistic regression model considered by \ctn{casella_bk}, given the O-ring data ${\bf D}$. The filled black circles represent $p_{mode}(t_i)$, where $T=t_i$ is the temperature in the $i$-th row of the O-ring data, $i=1,\ldots,23$. Thee distinct SFN-shaped functions of $T$, that approximate $p_{mode}(T)$ differently, i.e. are differently distant from $p_{mode}(T)$, are depicted: $p_{blue}(T)$ in the broken (blue in the e-version) lines, $p_{red}(T)$ in dotted (red in the e-version) lines and $p_{green}(T)$ in the long-dashed (green in the e-version) lines.}
\label{fig:ptemp}
\end{figure*}

Then $p_{mode}(T)$ is the variation in the failure probability with temperature that describes the measured data ${\bf D}$. We approximate $p_{mode}(T)$ with model function $p_k(T)$, where $k$ is a string-valued variable, $k=''red'', ``blue'', ``green''$, with the quality of the approximation parametrised by the constant mean square distance $\alpha_k$:
\begin{equation}
\alpha_k = \displaystyle{\frac{\sum\limits_{i=1}^{23}\left(p_{mode}(t_i) - p_k(t_i)\right)^2}{23}}.
\label{eqn:alpha}
\end{equation}
 The variation of failure probability with $T$, as displayed in Figure~\ref{fig:ptemp}, reminds us of the shape of a (scaled) folded-normal density function \ctp{folded}. This motivates us to choose a scaled-folded-normal functional form for $p_k(T)$, as follows.
\begin{equation}
p_k(T) = s_k \displaystyle{\left[\exp\left(-\frac{(T - m_k)^2}{2 v_k}\right) + \exp\left(-\frac{(T + m_k)^2}{2 v_k}\right)\right]},
\label{eqn:ourmodel}
\end{equation}
where the parameters of this function--the scaled-folded-normal (SFN) function--are: $S\in{\mathbb R}_{\geq 0}$, $M\in{\tau}\subset{\mathbb R}$ and $V\in{\mathbb R}_{\geq 0}$, which take values $s_k, m_k, v_k$ in the SFN-shaped variation $p_k(T)$ of failure probability with temperature. Thus, in the $k$-th model, the model parameter vector is $\btheta_k=(s_k, m_k, v_k)^T$, $k=``red'', ``blue'', ``green''$. Table~\ref{table:benchmark} includes the constant mean squared distance parameter, $\alpha_k$, that defines the SFN function $p_k(T)$, given $p_{mode}(T)$.  

We want to test for the null $H_0^{(k)}$, given the O-ring data. Here $H_0^{(k)}$ states that the measurable $Y$--measurements of which comprise ${\bf D}$--is distributed as Bernoulli with probability for a ``fail'' that is an SFN-shaped function of $T$, namely $p_k(T)$, that approximates $p_{mode}(T)$ s.t. the mean squared distance between these two functions computed at $t_1,\ldots,t_{23}$ is a constant $\alpha_k$, (presented in the 6-th column of Table~\ref{table:benchmark}). Then if at temperature $T=t$, the measurable is $Y=y$ (=1 or 0 for fail or not-fail, respectively), the $k$-th null is
\begin{eqnarray}
H_0^{(k)}: \Pr(Y=y)&=&\left(p_k(t)\right)^{y} \left(1 - p_k(t)\right)^{1-y}, \quad\mbox{where} \nonumber \\
p_k(T) \quad {\mbox{is an SFN function of $T$, s.t.}},& & 
\displaystyle{\frac{\sum\limits_{i=1}^{23}\left(p_{mode}(t_i) - p_k(t_i)\right)^2}{23}}=\alpha_k,
\label{eqn:ournull}
\end{eqnarray}
$k=''blue'', \:''red'',\:''green''$. Here the constant
$\alpha_{blue}=0.00005657,\alpha_{red}=0.001411,\alpha_{green}=0.01234$
and $t_i$ is the temperature in the $i$-th row of the O-ring data.
Thus, the $k$-th null is not sharp, for any $k$. By null $H_0^{(k)}$,
the observed temperature variation of O-ring failure rate is described
by $p_{k}(T)$, where ${p}_{k}(T)$ is known to be an approximation to
$p_{mode}(T)$ with the quality of the approximation parametrised by
the given distance $\alpha_{k}$ between them. Now, $p_{mode}(T)$
describes ${\bf D}$ well, as learnt by \ctn{casella_bk}. Thus, the
O-ring data is described approximately well by $p_{k}(T)$, where the
quality of such an approximation is given by how well $p_k(T)$
approximates $p_{mode}(T)$, i.e. how small $\alpha_k$ is. Thus, the
smaller the $\alpha_k$, the better does $p_k(T)$ describe the data
${\bf D}$, i.e. higher is the support in ${\bf D}$ for
$H_0^{(k)}$. Then we expect high support in ${\bf D}$ for
$H_0^{(blue)}$ as $\alpha_{blue}$ is small (smallest of the three
models considered).  On the other hand, owing to the higher value of
$\alpha_{red}$, support in ${\bf D}$ for $H_0^{(red)}$ is expected to
be less than for $H_0^{(blue)}$. Equally, support in ${\bf D}$ for
$H_0^{(green)}$ is expected to be least, as $p_{green}(T)$ is the
worst of the three approximations to $p_{mode}(T)$ (corroborated in
Figure~\ref{fig:ptemp}).

Values of $S, M, V$ that can define the SFN function $p_k(T)$ that
approximates $p_{mode}(T)$ according to given distance $\alpha_k$, are
tabulated in Table~\ref{table:benchmark} for each $k$. This table also
includes $\Pr(H_0^{(k)}\vert{\bf D})$, which is the support for the
$k$-th null in the measured O-ring data ${\bf D}$ that comprises
measured values of $Y$. The last column of this table gives the
logarithm of the ratio $\Omega_k$ of the marginalised posterior of
$\btheta_k$, given data ${\bf D}$ to the data ${\bf D}^/_k$ that is
generated from the $k$-th model of thermal variation in the O-ring
data (to be precise, ${\bf D}^/_k$ comprises 23 random numbers, the
$i$-th of which is sampled from a Bernoulli distribution with rate
$p_k(t_i)$, $i=1,\ldots, 23$).

\begin{table}
\caption{}
 \label{table:benchmark}
\begin{center}
{\footnotesize
\begin{tabular}{|c||c|c|c|c|c|c|c|}\hline
\hline
k& SFN function used & $s_k$ & $m_k$ & $v_k$ & $\alpha_k$ & $\Pr(H_0^{(k)}\vert{\bf D})$ & $\lg\left(\Omega_k\right)$\\
\hline
$red$ & ${p}_{red}(T)$ & 0.91 & 53.4 & 98.1 & 0.001411 & 0.8168 & -1.0814\\
$blue$ & ${p}_{blue}(T)$ & 0.97 & 51.7 & 99.0 & 0.00005657 & 1 & 2.8893\\
$green$ & ${p}_{green}(T)$ & 1.02 & 48.0 & 96.5 & 0.01234 & 0 & -8.5292\\
\hline
\end{tabular}
}
\vspace{-0.1in}
\end{center}
\end{table}

Here, the values of $\alpha_k$ are not arbitrarily chosen, but very much
motivated by aspects of this application. $p_{blue}(T)$ is the least
squares fit of an SFN-shaped function of $T$ to the sample $\{(t_i,
p_{mode}(t_i))\}_{i=1}^{23}$ taken from $p_{mode}(T)$ that is learnt
by \ctn{casella_bk} (the filled black circles in
Figure~\ref{fig:ptemp}); $p_{bue}(T)$ is depicted in this figure in
blue broken lines. The fit has a mean square error (MSE) of
$\alpha_{blue}$ of about 0.00005657. Figure~\ref{fig:ptemp} also includes the
SFN function ${p}_{red}(T)$ in 
dotted (red) lines. 
${p}_{red}(T)$ is only a moderately good fit with an
MSE of about 0.001411 (=$\alpha_{red}$). 
This SFN function $p_{red}(T)$ is parametrised by the modal values of $S$, $M$ and
$V$ that are learnt using data ${\bf D}$ in an MCMC-based inference scheme. To achieve the modal values of $S,M,V$,
we model $p(T)$ as an SFN function with unknown
parameters $S,M,V$, so that the likelihood is rendered as in the
RHS of Equation~\ref{eqn:logis}, except now $p_i$ is the value of the SFN function $p(T)$ computed at $T=t_i$. Using this likelihood and flat priors on all three unknown parameters,
we generate posterior samples from $\pi(S,M, V\vert {\bf D})$ using
Random-Walk Metropolis-Hastings. Let us refer to this MCMC chain as
``Chain~I'' for future reference.  The trace of this joint posterior
probability in this chain is shown in Figure~\ref{fig:oring_res1} in
the solid black line. The marginals of $S$, $M$ and $V$ are shown in
Figure~\ref{fig:oring_res2}. So when the modal values of these
marginals are employed as $s_{red}, m_{red}$ and $v_{red}$ (see
columns 3,4,5 of Table~\ref{table:benchmark}), in an SFN function of
$T$ (Equation~\ref{eqn:ourmodel}), $p_{red}(T)$
results, which is $\alpha_{red}$ distance away from $p_{mode}(T)$. $p_{green}(T)$ is constructed by choosing a value of $S,M,V$
each from the tails of their respective marginals learnt using ${\bf
  D}$ (Figure~\ref{fig:oring_res2}). $p_{green}$ is a bad
approximation of $p_{mode}(T)$, as parametrised by a high
$\alpha_{green}$ (of about 0.01234).

\begin{figure*}[!t]
     \begin{center}
  {
\includegraphics[height=3.5cm]{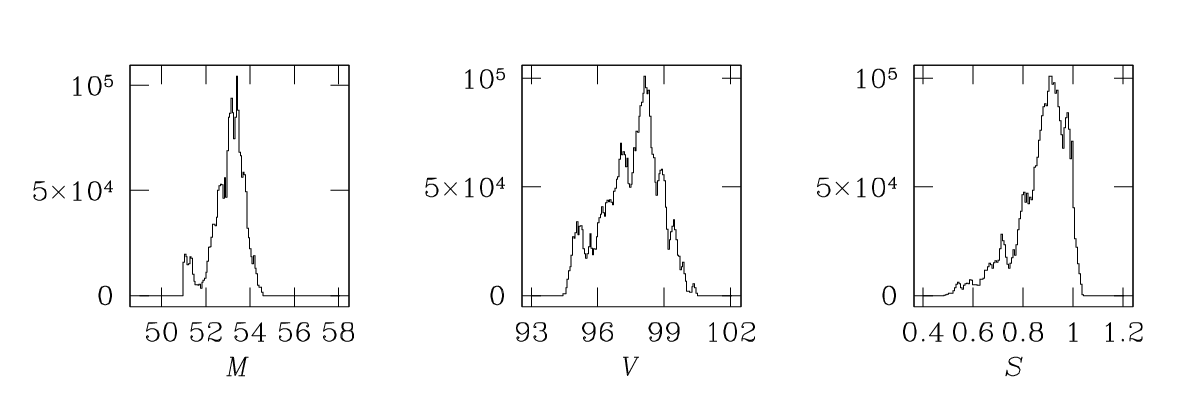} 
}
\end{center}
\vspace{-1cm}
\caption{Panels show marginals of the unknown parameters $S,M,V$ that parametrise an SFN function $p(T)$ that models the variation of failure probability with temperature $T$. These marginals are learnt using an MCMC chain, with the O-ring data. }
\label{fig:oring_res2}
\end{figure*}

\begin{figure*}[!t]
     \begin{center}
  {
\includegraphics[height=8.5cm]{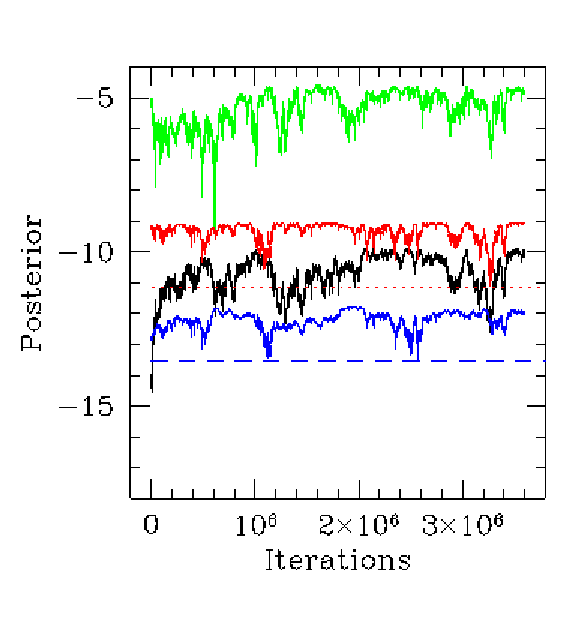} 
}
\end{center}
\vspace{-.5cm}
\caption{ In solid black: trace of the joint posterior probability
  density $\pi(S,V,M\vert{\bf D})$ of the unknown model parameters
  $S,M, V$, given measured data ${\bf D}$, from the MCMC chain
  Chain~I. In broken (blue) lines: trace of the posterior of $S,V,M$
  given generated data ${\bf D}^/_{blue}$ that is randomly sampled
  from a Bernoulli distribution with rate given by the SFN function
  $p_{blue}(T)$. This chain corresponds to the lowest posterior values
  amongst the four chains shown here. $\pi^{(min)}(S,V,M\vert{\bf
    D}^/_{blue})$ is depicted in the broken (blue) lines. In dotted
(red) lines: trace of $\pi(S,V,M\vert{\bf D}^/_{red})$ where ${\bf
  D}^/_{red}$ is generated using $p_{red}(T)$ as the variation in
failure probability with $T$; minimum of this posterior is shown in
(red) dots. In (green) long dashes: trace of $\pi(S,V,M\vert{\bf
  D}^/_{green})$ where ${\bf D}^/_{green}$ is generated using
$p_{green}(T)$. This chain occurs at the highest posterior density
values out of the four chains shown here.}
\label{fig:oring_res1}
\end{figure*}

The test is implemented using the following steps.
\begin{enumerate}
\item We consider the measurable $Y$ to be a Bernoulli variate with rate
  parameter that varies with temperature as $p(T)$--modelled as an SFN
  function with unknown model parameter vector $\btheta=(S, M,V)^T$. We perform Bayesian
  learning of these parameters given the measured data ${\bf D}$, in
  ``Chain~I''. $\pi(\btheta\vert{\bf D})$ is
  shown in Figure~\ref{fig:oring_res1} in the solid black line. 
\item We identify the benchmark model ${\cal M}_k$ in which the $k$-th null is true, $k=''red'', ``blue'', ``green''$. Then in model ${\cal M}_k$, the variation of failure probability with temperature is an SFN-shaped function $p_k(T)$, s.t. the mean squared distance between itself and $p_{mode}(T)$, computed at the temperature values in each row of the O-ring data, is $\alpha_k$. Such a function $p_k(T)$ is achieved using $\btheta_k$ that is given in Table~\ref{table:benchmark}. Then we attain the generated data ${\bf D}^/_k$ by selecting a random Bernoulli variate with rate given by this $p_k(T)$. We then run an MCMC chain with ${\bf D}^/_k$, to obtain samples from $\pi(\btheta\vert{\bf D^/}_k)$. (This chain is of course different from ``Chain~I'' that is run with data ${\bf D}$). We employ this chain to identify the minimum value of $\pi(\btheta\vert {\bf D^/}_k)$. Trace of the posterior in this chain is shown in Figure~\ref{fig:oring_res1} in (colour $k$ in the electronic version) dashed lines for $k=''blue''$, dotted lines for $k=''red''$, broad-dashed lines for $k=''green''$.  The minimum posterior in the post-burnin part of the chain is also presented in the figure as a horizontal line in the corresponding line-type.
\item Next we identify the sub-space ${\cal T}_{{\cal M}_k}({\bf D})$
  that is the native space of those model parameter vectors, for which
  $\pi(\btheta\vert {\bf D})$ equals or exceeds the minimum posterior
  attained under the $k$-th null, i.e. when $p_{mode}(T)$ is
  approximated by $p_k(T)$, within a distance parameter of
  $\alpha_k$. Once we identify this sub-space, we then need to compute
  $\Pr(\btheta\in T_{{\cal M}_k}\vert{\bf D}) =
  \displaystyle{\int\limits_{\btheta\in {\cal T}_{{\cal M}_k}({\bf
        D})} \pi(\btheta\vert{\bf D})d\btheta}$. However, we avoid the
  computation of this integral, and instead approximate the
  probability of membership in this sub-space via a
  simple case-counting scheme. Thus, we identify the number $P_k$ out
  of the total of $Q_k$ $\btheta$ samples that are generated in the
  MCMC chain ``Chain~I'', run with measured data ${\bf D}$, for which
  posterior probability exceeds, or is equal to
  $\pi^{(min)}(\btheta\vert {\bf D^/}_k)$. Then, $\Pr(\btheta\in
  T_{{\cal M}_k}\vert{\bf D})$ is approximated by
  $\displaystyle{\frac{P_k}{Q_k}}$. Then by
  Equation~\ref{eqn:support}, the probability of the $k$-th null
  conditional on the measured data, is $\Pr(\btheta\in T_{{\cal
      M}_k}\vert{\bf D})\approx \displaystyle{\frac{P_k}{Q_k}}$. This
  is tabulated in the 7-th column of Table~\ref{table:benchmark} for
  each $k=''red'', ``blue'', ``green''$. The 8-th column contains the
  logarithm of the odds ratio $\Omega_k$ discussed in
  Equation~\ref{eqn:oddsratio}.
\end{enumerate}

As said above in the paragraph following Equation~\ref{eqn:ournull}, we
expect high support in ${\bf D}$ for $H_0^{(blue)}$. In fact, in the
chain run with generated data ${\bf D}^/_{blue}$,
$\pi^{(min)}(\btheta\vert {\bf D^/}_{blue})$ is about -13.55, which is
lower than $\pi(\btheta\vert {\bf D})$ obtained for all $\theta$
samples generated in Chain~I (in solid black line in
Figure~\ref{fig:oring_res1}),
i.e. $\displaystyle{\frac{P_{blue}}{Q_{blue}}}\approx \Pr(\btheta\in
T_{{\cal M}_{blue}}\vert{\bf D})=\Pr(H_0^{(blue)}\vert {\bf
  D})=1$--the highest support possible in the measured data.
Compared to $H_0^{(blue)}$, support in ${\bf D}$ for $H_0^{(red)}$ is
expected to be less. Indeed we find that
$\displaystyle{\frac{P_{red}}{Q_{red}}}\approx 0.8168$ or
equivalently, $\Pr(\btheta\in T_{{\cal M}_{red}}\vert{\bf
  D})=\Pr(H_0^{(red)}\vert{\bf D})$ is about 0.8168. Here
$\pi^{(min)}(\btheta\vert {\bf D^/}_{red})\approx -11.14$.
For the crudest (out of the three models) approximation for $p_{mode}(T)$, 
in the chain run with generated data ${\bf D}^/_{green}$, the minimum
posterior probability exceeds the posterior achieved for every $\btheta$
sample generated in Chain~I that is run with measured data ${\bf
  D}$. Then fraction of these samples for which posterior exceeds of
equals posterior achieved in chain run with generated data, is 0, i.e.
$\displaystyle{\frac{P_{green}}{Q_{green}}}=0$ implying
$\Pr(H_0^{(green)}\vert{\bf D})=0$.

As in this application we are not comparing support in one data set for a given null, to support in another data for a different null, we could have computed the support in the measured O-ring data ${\bf D}$, for the $k$-th null, using the ratio of the marginalised posterior given ${\bf D}$ to that given ${\bf D}^/$ that is defined in Equation~\ref{eqn:oddsratio} as $\Omega_k$. In this example, we can perform posterior computation given measured and generated data; $\int\limits_{\btheta} \pi(\btheta\vert{\bf D})$ is about $5.4\times 10^{-10}$, and 
$\int\limits_{\btheta} \pi(\btheta\vert{\bf D}^/_k)$ is about $2.7\times 10^{-6}$, $1.6\times 10^{-9}$, $3.0\times 10^{-11}$, for $k=''green'', ``red'', ``blue''$, so that support in ${\bf D}$ for the $k$-th model as in $\log(\Omega_k)$, is about -8.53, -1.08, 2.89 for $k=''green'', ``red'', ``blue''$ respectively (see Table~\ref{table:benchmark}). 

\section{Implementation of the new test to the galactic application}
\label{sec:implementation}
\noindent 
Following Section~\ref{sec:new}, 
we implement the new
test by
finding the minimum
posterior achieved under the null, in order to identify the sub-space
${\cal T}_{{\cal M}_i}({\bf D}_i)$, and then proceed to compute the probability of
the null given data ${\bf D}_i$, as the probability that
$\btheta_i\in{\cal T}_{{\cal M}_i}({\bf D}_i)$.

Let the model parameter vector that minimises the posterior
probability density under the null, be referred to as $\btheta_{i}^{(min)}$.

\subsection{Identification of posterior-optimising model parameter vector, under the null}
\label{sec:1ststep}
\noindent
In order to identify the vector, $\btheta_{i}^{(min)}$, the
following scheme is used, where the scheme below is expressed in the
paradigm of the Bayesian method in which the discretised state space
density vector ${\bPsi}^{(i)}$ and the discretised gravitational mass
density vector $\brho^{(i)}$ are learnt given the measured data ${\bf
  D}_i$, under the assumption that the state space $pdf$ is isotropic
(see Section~\ref{sec:null}). The benchmark model ${\cal M}_i$ is such,
that under it, the state space $pdf$ is an isotropic function of the
location $\bX$ and velocity $\bV$ of a galactic particle, i.e. the null
$H_0^{(i)}$ is true in model ${\cal M}_i$.
\begin{itemize}
\item We perform inference on $\btheta_{i}$ given
  measured data ${\bf D}_i$, with
  Metropolis-Hastings. During this inference, let the state space vector in
  the $c$-th iteration be $\btheta_{i}^{(c)}$, $c=1,\ldots,N_0$, where
  the chain is $N_0$ steps long. Upon convergence, the unknown
  $\btheta_i$, i.e. ${\bPsi}_{i}$ and ${\brho}_{i}$ in our application, are learnt within
  95$\%$ HPD credible regions. From a given chain, we identify the
  modal parameter
  vector $\btheta_i^{(M)}:=(\Psi_1^{(M,i)},\ldots,\Psi_{N_E}^{(M,i)},\rho_1^{(M,i)},
  \rho_{N_x}^{(M,i)})^T$, corresponding to the mode of the posterior
  density $\pi(\btheta_{i}\vert{\bf D}_i)$.
\item We learn the discretised state space density ${\bPsi}^{(M,i)}$
  and gravitational mass density $\brho^{(M,i)}$ given ${\bf D}_i$, in
  the aforementioned Bayesian method, where the learnt state space
  density is isotropic by construct, (since isotropy of the state space
  density is the basic underlying assumption of the Bayesian method). From this
  learnt isotropic $pdf$, at the learnt $\brho^{(M,i)}$, we
  simulate an $N_{data}^{(i)}$-sized data set of the observed
  variables $X_1$, $X_2$ and $V_3$. Let this generated data
  set be \\
$${\bf D}_{i}^{(gen)}:=\{(x_{1,\:gen}^{(k)}, x_{2,\:gen}^{(k)},
  v_{3,\:gen}^{(k)})\}_{k=1}^{N_{data}^{(i)}},$$
where the size of ${\bf D}_i$ is $N_{data}^{(i)}$. 
\item Importantly, generated data ${\bf D}_{i}^{(gen)}$ is simulated from an isotropic state space function (the discretised form of which is) ${\bPsi}^{(M,i)}$, at ${\brho}^{(M,i)}$, using rejection
  sampling, according to the following algorithm.
\begin{enumerate}
\item We solve for the function $\Phi(X)$ that relates to the sought
  unknown $\rho(X)$ via the
  Poisson equation: $\nabla^2\Phi(X)=-4\pi G \rho(X)$, where
  $X:=\parallel\bX\parallel$. The relevance of $\Phi(X)$ is that it is
  part of the function $E(X,V)$ ($=\Phi(X)+\eta(V)$) that was
  introduced in Section~\ref{sec:null}, where the function
  $E(\cdot,\cdot)$ forms the argument of state space density:
  $\Psi_i(E(X,V))$. By its dependence on $X$ and $V$, (via $E(X,V)$),
  this model of the state space $pdf$ is an isotropic function of
  $\bX$ and $\bV$ (see Section~\ref{sec:null}). Then isotropic state space
  $pdf$ bears the form $\Psi_i(E(X,V))$ orequivalently, the form $\Psi_i(\Phi(X),\eta(V))$ which is again equivalent in form to
  $\Psi_i(\rho(X), \eta(V))$, by invoking Poisson equation. In this
  way, the discretised version $\brho$, of $\rho(X)$, can be embedded
  into the argument of the state space density that is modelled as
  isotropic; $\brho$ thereby enters the likelihood of the unknowns
  given the data, thus allowing for inference on the unknown $\brho$.
\item In our application, $E(X,V)$ is identified with the total energy
  of a galactic particle, with $\Phi(X)$ the potential and
  $\eta(V)=V^2/2$ identified with the kinetic energy. In fact in our
  application, $\Phi(X)\leq 0$ for $\rho(X)\geq 0$ and the minimum
  value of $\Phi(X)$ is $\Phi(0)$.  We consider only those galactic
  particles that are bound to the galaxy; the energy of any such bound
  particle is negative. Thus, in this application, $E(X,V)$ can at
  most approach 0, and at least be $\Phi(0)$.  Thus, the value
  $\epsilon$ of $E(X,V)$ normalised by $\Phi(0)$, lies in (0,1].
\item Since the value of $E(X,V)\:(=\Phi(X)+ V^2/2)$
is minimally $\Phi(0)$, and maximally approaches 0, the range of values of $V$ is $[-\sqrt{-2\Phi(0)}, \sqrt{-2\Phi(0)}]$. 
\item We discretise $\rho(X)$ by discretising the range that $x$ lies in,
  and discretise $\Psi(E)$ by discretising the range that
  $\epsilon$ lies in. Thus, $\rho_p=\rho(x)$ if $x\in[(p-1)\delta, p\delta)$ and $\Psi_t=\Psi(\epsilon)$ if $\epsilon\in[(t-1)\delta_E, t\delta_E)$, for $p=1,\ldots,N_x$, $t=1,\ldots,N_E$. (Though we use uniform binning in this application--with constant bin widths $\delta>0$ and $\delta_E> 0$--other forms of discretisation can be potentially implemented within this scheme).
\item We compute $\Phi(x)$ via $M(x)$ where
  $\displaystyle{\Phi(x)=\frac{-GM(x)}{x}}$ with
  $M(x)=\displaystyle{\int_{s=0}^x 4\pi\rho(s)s^2 ds}$ and $G$ is a
  known (Universal Gravitational) constant. For computational ease we discretise this integral, to define
\begin{eqnarray}
M(x) &=& \displaystyle{\sum_{q=1}^p \frac{4\pi}{3}[q^3\delta^3 - 
(q-1)^3\delta^3]\rho_q + \frac{4\pi}{3}[x^3 -p^3\delta^3]\rho_{p+1}},\quad
{\textrm{for}}\quad x\in[p\delta, (p+1)\delta),\nonumber \\
M(x) &=& \displaystyle{\sum_{q=1}^{N_x} \frac{4\pi}{3}[q^3\delta^3 - 
(q-1)^3\delta^3]\rho_q} \quad
{\textrm{for}} \quad x \geq N_x\delta, \nonumber \\
M(x) &=& \displaystyle{\frac{4\pi}{3}[x^3]\rho_1} 
\quad {\textrm{for}} \quad 0\leq x \leq \delta. \nonumber \\
\label{eqn:phi}
\end{eqnarray}
Here $N_x\delta$ is the maximum radius to
which data are available and $\rho_q$ is the gravitational mass
density in the $q$-th radial bin. This defines $\Phi(x)$ for any $x\geq
  0$, given the identified ${\brho}^{(M,i)}$.
\item Next, we sample $\epsilon$, i.e. the value of $E(\cdot,\cdot)$ normalised by $\Phi(0)$. As $\epsilon\in(0,1]$, we choose $\epsilon$ randomly from
  ${\cal U}[0,1]$, where ${\cal U}[a,b]$ is the uniform
  distribution over the range $[a,b]$, $a,b\in{\mathbb
    R}$. Let the sampled $\epsilon$ be such that it lies in the $t$-th
  energy bin, i.e. $\epsilon\in[(t-1)\delta_E, t\delta_E]$,
  $t=1,\ldots,N_E$; let the $t$-th
  component of ${\bPsi}^{(M,i)}$ be $\Psi_t^{(M,i)}$.

\item The 3 components of the location vector are continuous in $[-N_x\delta,
      N_x\delta]$. So we sample, $X_1, X_2, X_3\sim{\cal U}[-N_x\delta,
      N_x\delta]$ and using these sampled values
  $x_1,x_2,x_3$, obtain the value of $\parallel\bx\parallel\equiv x =
  \sqrt{x_1^2 + x_2^2 + x_3^2}$. Let $x$ be such that it lies in the
  $q$-th radial bin, i.e. $x\in[(q-1)\delta, q\delta]$,
  $q=1,\ldots,N_{x}$. For this chosen $x$, we then compute $\Phi(x)$ using $M(x)$ from Equation~\ref{eqn:phi} and the definition $\displaystyle{\Phi(x)=\frac{-GM(x)}{x}}$. We normalise $\Phi(x)$ by $\Phi(0)$, so that $\Phi(x)$ now lives in the range $(0,1]$.

\item Check if the chosen $\epsilon > \Phi(x)$. If not, go back to
  step number 6. If yes, then recall that the components of the velocity vector, $V_1$, $V_2$, $V_3$ is each
  continuous in $[-\sqrt{-2\Phi(0)},\sqrt{-2\Phi(0)}]$, to suggest that $V_1, V_2, V_3$ be each sampled as $V_1, V_2, V_3\sim{\cal
    U}[-\sqrt{-2\Phi(0)},\sqrt{-2\Phi(0)}]$. So we draw $v_1,v_2,v_3$
  individually from this uniform distribution.

\item In this step, we sample from $\Psi_t^{(M,i)}$ using rejection
  sampling. Here the chosen $\epsilon$ is in the $t$-th energy-bin so
  that $\Psi_t^{(M,i)}$ is the value of the state space $pdf$ in our discretised
  model. The rejection sampling is done by checking
  if $\displaystyle{\frac{\Psi_t^{(M,i)}}{g(\epsilon)}} > u$ or not, where
    $u$ is a random number in $[0,1]$, $u\sim{\cal U}[0,1]$.  Here
    $g(\epsilon)$ is the proposal density function that is chosen to
    envelope over $\Psi(\epsilon)$, $\forall \epsilon$, and is defined as
    $g(\epsilon) = 1.05 \forall \epsilon$. This is an adequate choice
    because the state space $pdf$ $\Psi(\epsilon)$ is normalised to be in
    $(0,1]$. If the above inequality holds, we allow an integer-valued
    flag, $\gamma$, an increment of 1 and accept the values $x_1, x_2$
    and $v_3$ as chosen in steps 7 and 8 respectively, as the
    $\gamma$-th data point in ${\bf D}_i^{(gen)}$. We iterate over points 4
    to 9, until $\gamma$ equals $N_{data}^{(i)}$.
\end{enumerate}

\item Now that we have discussed the algorithm used to sample the generated data
  ${\bf D}_i^{(gen)}$, in order to estimate $\btheta_i$ using this
  generated data, we start a new MCMC chain. We remind ourselves that
  unlike the measured data ${\bf D}_i$ that may live in an anisotropic
  state space, the generated data ${\bf D}_i^{(gen)}$ is sampled from
  an isotropic state space density (rather its discretised form
  $\bPsi_i)$, i.e. posterior of $\btheta_i$ given data ${\bf
    D}_i^{(gen)}$ is the posterior when the null is true. Post burn-in,
  samples of $\btheta_{i}$ vectors generated in each iteration are
  recorded. In this recorded sample of values of $\btheta_{i}$, that
  which minimises the posterior density
  $[\Psi_1^{(i)},\ldots,\Psi_{N_E}^{(i)},\rho_1^{(i)},\ldots,\rho_{N_x}^{(i)}\vert{\bf
      D}_i^{(gen)}]$, is the posterior-minimising parameter in the
  benchmark model ${\cal M}_i$:
\begin{equation}
  \btheta_i^{(min)}:=(\Psi_1^{(i,min)},\ldots,\Psi_{N_E}^{(i,min)},\rho_1^{(i,min)},\ldots,\rho_{N_x}^{(i,min)})^T.
\end{equation}
Let the minimum posterior of $\btheta$ given the generated data be $\pi^{(min)}(\btheta_i\vert {\bf D}_i^{(gen)})$.

\end{itemize}

\subsection{Probability of membership in subspace ${\cal T}_{{\cal M}_i}({\bf D}_i)$}
\noindent
We need to identify the sub-space ${\cal T}_{{\cal M}_i}({\bf D}_i)$ in which
live model parameter vectors, the posterior of which equals or exceeds
the minimal posterior probability density attained under the null, i.e. $\pi^{(min)}(\btheta_i\vert
{\bf D}_i^{(gen)})$. We are required to integrate the posterior
probability density of $\btheta_i$ given measured data ${\bf D}_i$,
over all such values of $\btheta_i$ that live in the subspace ${\cal
  T}_{{\cal M}_i}({\bf D}_i)$, i.e. compute
$\displaystyle{\int\limits_{\btheta_i\in {\cal T}_{{\cal M}_i}({\bf D}_i)}
  \pi(\btheta_i\vert{\bf D}_i)d\btheta_i}$. This integral is then
equal to $\Pr(\btheta\in T_{{\cal M}_i}\vert{\bf D}_i)$.

Thus, in this approach, it is possible to implement
$\Pr(H_0^{(i)}\vert {\bf D}_i)$, even in a high-dimensional state space, by approximating this
probability of membership of the model parameter vector $\btheta_i$ in 
${\cal T}_{{\cal M}_i}({\bf D}_i)$, with a case-counting scheme. In other
words, we compute the proportion of the model parameter vectors for
which $\pi^{(min)}(\btheta_i\vert {\bf D}_i^{(gen)}) \leq
\pi(\btheta_i\vert {\bf D}_i)$, as recovered in the post-burnin stage
of chains run with measured data ${\bf D}_i$.

Thus, let there be a total of $Q_i$ number of samples of $\btheta_i$
vectors recovered in the post-burnin stage in chains run with
measured data ${\bf D}_i$. Out of these, let $P_i$ number of $\btheta_i$
vectors be such that $\pi^{(min)}(\btheta_i\vert {\bf
  D}_i^{(gen)}) \leq \pi(\btheta_i\vert {\bf D}_i)$. Here, $Q_i,\;
P_i\in{\mathbb Z}_+$, $P_i\leq Q_i$. Then the fraction $P_i/Q_i$ is an
approximation to the probability that $\btheta_i\in {\cal T}_{\cal M_{i}}({\bf
  D}_i)$. Then recalling Equation~\ref{eqn:support},
we state that 
\begin{equation}
\Pr({H_0^{(i)}}\vert{\bf D}_i) = \displaystyle{\frac{P_i}{Q_i}},
\end{equation} 
$i$=1,2. 


\section{Testing with synthetic galactic data}
\label{sec:simulated}
\noindent
In this section, we implement this new test to find the
probability of the null (that the state space of a toy galaxy is isotropic),
given the (simulated) data at hand. For this simulation exercise, we use synthetic data that is sampled from chosen state space density
models, constructed to simulate real galactic state space density
functions. To be precise, we sample data sets ${\bf D}_{A}$ and ${\bf
  D}_{B}$ from two chosen state space density functions
$f_A^{(True)}(\bX,\bV)$ and $f_B^{(True)}(\bX,\bV)$
respectively, that are anisotropic to different extents, as
parametrised by an anisotropy parameter that we discuss below. We
realise that a state space density that is a function of $\bX$ and
$\bV$ via a function such as $E(X, V)$, is an isotropic function of vectors
$\bX$ and $\bV$. On the other hand, a density function that depends on
$\bX$ and $\bV$ via any form of these vectors, other than their
$L^2$-norm, is not an isotropic function of $\bX$ and $\bV$. 

The model state space $pdf$ that we sample the synthetic data ${\bf D}_{A}$ and ${\bf
  D}_{B}$ from, are
\begin{eqnarray}
f_{\cdot}^{(True)}({\bf x},{\bf v}) &=& \displaystyle{\frac{1}{\sqrt{2\pi}\sigma}\exp\left(\frac{\epsilon(x, v)}{2\sigma^2}\right)\exp\left(-\frac{[P(\bx,\bv)]^2}{r_a^2 \sigma^2}\right)}
\label{eqn:form}
\\
{\mbox{where}}\quad \epsilon(x,v) &=& \displaystyle{\frac{{{v}^2}/{2} +\Phi(x)}{\Phi_0}},  \\
{\mbox{and}}\quad [P(\bx,\bv)]^2 &=& (x_2 v_3 - x_3 v_2)^2+ (x_3 v_1 - x_1 v_3)^2 + (x_1 v_2 - x_2 v_1)^2,
\end{eqnarray}
and $r_a$ and $\sigma$ are parameters of this density.
The first exponential term in the RHS of Equation~\ref{eqn:form} manifests the purely isotropic dependence on $\bX$ and $\bV$, while the second exponential term manifests dependence on $\bX$ and $\bV$ via a form that is different from the $L^2$-norm of these vectors, i.e. this second exponential term manifests anisotropic dependence on $\bX$ and $\bV$. Thus, the chosen state space density functions of the type in Equation~\ref{eqn:form}, are anisotropic in general, with the strength of the (anisotropic) second exponential factor on the RHS of Equation~\ref{eqn:form}, parametrised by the parameter $r_a$; the bigger is the value of $r_a$, higher is the relative amplitude of the anisotropic factor to the isotropic factor (that is parametrised only by $\sigma$). Equally, for $r_a$ approaching 0, the constructed state space $pdf$ in Equation~\ref{eqn:form} approaches an isotropic form. The parameter $r_a$ is then the anisotropy scale length. It is measured in the astronomical unit of length on galactic scales: ``kiloparsec'', abbreviated to ``kpc''.

We choose $f_A^{(True)}(\bX,\bV)$ to be more anisotropic than
$f_B^{(True)}(\bX,\bV)$ by choosing $r_a$=4 kpc and $r_a$=0.2
kpc in the two models respectively. In every other way, inputs to $f_A(\bX,\bV)$ and
$f_B(\bX,\bV)$ are identical. We choose $\sigma=220$, in units of
km s$^{-1}$. To define $E(X,V)$ and thereby its value $\epsilon$ in
Equation~\ref{eqn:form}, we need to choose the form of $\Phi(X)$. We
construct this to be
\begin{equation}
\Phi(x) = \displaystyle{-\frac{G M_0}{\sqrt{r_c^2 +x^2}}},
\end{equation}
where we chose the parameters to be $M_0=4\times 10^{11}$ times the
mass of the Sun or ``$M_\odot$'' (astronomical unit of mass on galactic
scales) and $r_c$=8 kpc. $G$ is a known physical constant, (the
Universal Gravitational constant). 

Having constructed $f_A^{(True)}(\bX,\bV)$ and
$f_B^{(True)}(\bX,\bV)$, we simulate data ${\bf D}_{A}$ and ${\bf
  D}_{B}$ respectively from these state space densities, where each
data set contains information on $X_1$, $X_2$ and $V_3$. Size of
${\bf D}_A$ is 710 while size of ${\bf D}_B$ is 135. The sampled $V_3$
data is chosen to be characterised by Gaussian noise $\sim{\cal N}(0,
20^2)$ which is typical of real-life galaxies that are nearby
\ctp{douglas_07}. 

The $i$-th null states that the data ${\bf D}_i$ is sampled from an isotropic state space density  $f_i(\bX,\bV)$ for $i=A,B$, i.e. $f_i(\bX,\bV)=\Psi_i(E(X,V))$, $\Psi_i(\cdot)\geq 0$, where $X\in{\cal X}\subseteq{\mathbb R}_{\geq 0}$ and $V\in{\cal V}\subseteq{\mathbb R}_{\geq 0}$. To condense, 
\begin{equation}
H_0^{(i)}:f_i(\bX,\bV)=\Psi_i(E(X,V)), \:\:\Psi_i(\cdot)\geq 0,
\label{eqn:null_syn}
\end{equation}
 for $i=A,B$. When the null is true, the state space $pdf$ is an isotropic function of $\bX$ and $\bV$.
As discussed above for our application, the intractability of the more
complex model (anisotropic state space $pdf$) compels us to learn the model parameter $\btheta_i$
only under the null model, i.e. by assuming the state space to be
isotropic. The model parameter vector for $i=A$ is
$\btheta_A=(\Psi_1^{(A)},\ldots,\Psi_{N_E}^{(A)}, \rho_1^{(A)},
\ldots, \rho_{N_x}^{(A)})^T$ is learnt using data ${\bf D}_A$ under the assumption that the galactic state space is isotropic,
where $\brho_A:=(\rho_1^{(A)}, \ldots, \rho_{N_x}^{(A)})^T$ and
${\bPsi}_A:=(\Psi_1^{(A)}, \ldots, \Psi_{N_E}^{(A)})^T$. Similarly, we
define $\btheta_B$, $\brho_B$, $\bPsi_B$, learnt using data ${\bf
  D}_B$, while assuming an isotropic galactic state space.

In Figure~\ref{fig:sim_res} we present the posterior probability density
$\pi(\btheta_A\vert{\bf D}_A)$ (right panel) and
$\pi(\btheta_B\vert{\bf D}_B)$ (left panel), in grey (or red in the
electronic version). The posterior probability density attained under
the null, i.e. computed given the generated data, is shown in black in each case: $\pi(\btheta_A\vert{\bf D}_A^{(gen)})$ in the right
and  $\pi(\btheta_B\vert{\bf D}_B^{(gen)})$ in the left
    panel. We recall that the generated data sets ${\bf D}_i^{(gen)}$ 
are generated using rejection sampling from $\Psi_i(E)$--or rather its discretised version $\bPsi_i$ that is learnt using available measurements ${\bf D}_i$--at the estimated $\brho_i$. See Section~\ref{sec:implementation} for details of implementation of this rejection sampling.   

It is clear that for the case of the more anisotropic true state space density, i.e. for case $A$, the posterior probability density of the model parameter vector falls below the minimal value of the posterior under the null, i.e. $\pi(\btheta_A\vert{\bf D}_A) < \pi^{(min)}(\btheta_A\vert{\bf D}_A^{(gen)}),\:\forall\btheta_A$, implying that the sub-space ${\cal T}_{{\cal M}_A}({\bf D}_A)$ is empty. It then follows that $\Pr(H_0^{(A)}\vert{\bf D}_A)=0$, so that we reject null $H_0^{(A)}$ with 100$\%$ probability. In other words, the hypothesis that the data ${\bf D}_A$ is sampled from an isotropic state space density is rejected at probability of 1. This is indeed what we expect given that the true density $f_A^{(True)}(\bX,\bV)$ that ${\bf D}_A$ is sampled from is chosen to be strongly anisotropic.

For the case of the less anisotropic true state space density,
i.e. for case $B$, in the post-burnin part of the chain (beyond the
600,000-th iteration; in black in Figure~\ref{fig:sim_res}),
$\pi^{(min)}(\btheta_B\vert{\bf D}_B^{(gen)})$ is depicted in the
solid black line. There are multiple values of $\pi(\btheta_B\vert{\bf
  D}_B)$ that exceed this minimal posterior achieved under the
null. In fact, in the post-burnin stage of the chain run with data
${\bf D}_B$, $\pi(\btheta_B\vert{\bf D}_B) \geq
\pi^{(min)}(\btheta_B\vert{\bf D}_B^{(gen)})$ for 83,780 samples of
$\btheta_B$ where there are 200,000 iterations, post-burnin in the
chain. Thus, for this case, $\Pr({\cal T}_{{\cal M}_B}\vert{\bf D}_B)=
\displaystyle{\frac{87650}{200000}}\approx 0.5394$, i.e. the support
against the null $H_0^{(B)}$ is 1-0.5394= 0.4606. Thus, the hypothesis
that the data ${\bf D}_B$ is sampled from an isotropic state space
density is rejected at probability 0.4606, given data ${\bf D}_B$.

This corroborates the strength of our test as we chose to sample data
${\bf D}_B$ from the true state space density $\Psi_B(\bX,\bV)$ that
is constructed as mildly anisotropic, compared to the
strongly anisotropic true density $\Psi_A(\bX,\bV)$ that data ${\bf
  D}_A$ is sampled from.

\begin{figure*}[!t]
     \begin{center}
  {
\includegraphics[height=8.5cm]{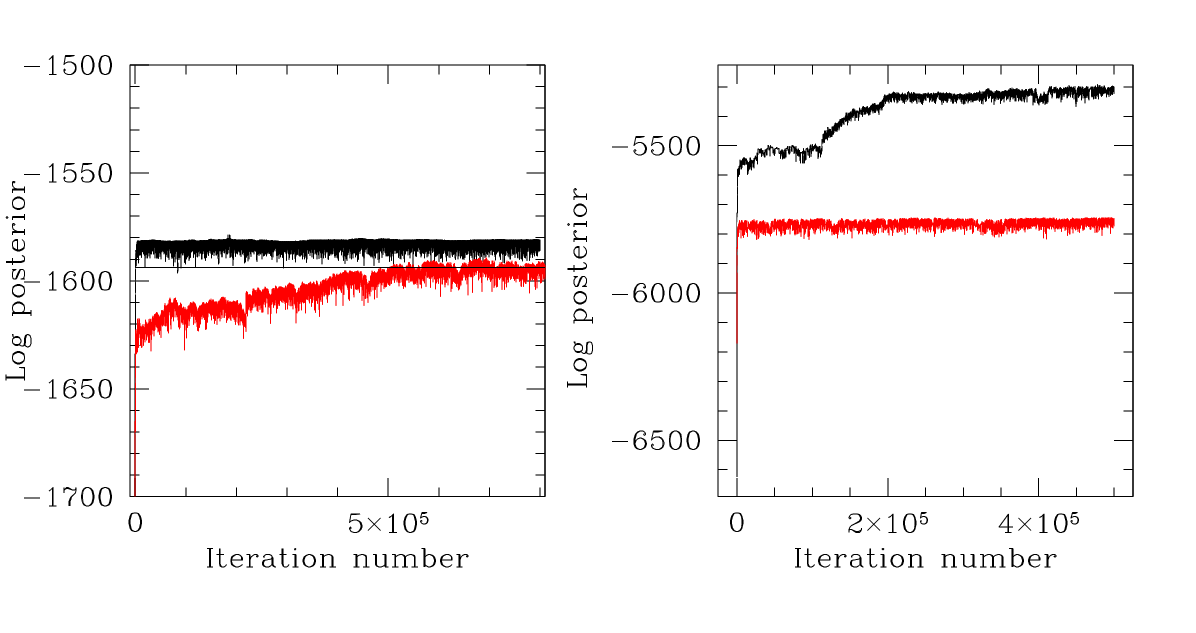} 
}
\end{center}
\vspace{-1cm}
\caption{Figure showing log of the posterior probability density
  $\pi({\btheta}_A\vert{\bf D}_A)$ (right) and
  $\pi({\btheta}_B\vert{\bf D}_{B})$ (left), in grey (or red in the
  electronic version), for chains that were run for 8$\times$10$^5$
  and 5$\times$10$^5$ iterations respectively. The log of the
  posterior probability density of $\btheta_A$ and $\btheta_B$, given
  generated data ${\bf D}_{A}^{(gen)}$ and ${\bf D}_{B}^{(gen)}$
  respectively, represent the posterior densities of the model
  parameters in the benchmark models in which the null is true; the
  traces of these posteriori are shown in black in the right and left
  panels. Here simulated data set ${\bf D}_A$ is about 5.3 times bigger in size than data ${\bf D}_B$. ${\bf D}_A$ is sampled from a true
  state space density that is constructed as strongly anisotropic, as
  distinguished from the mildly anisotropic true state space density
  that simulated data ${\bf D}_B$ is sampled from. In the right panel,
  the minimum value of the posterior when the null $H_0^{(A)}$ is true, is in
  excess of the posterior $\pi(\btheta_A\vert{\bf D}_A)$ at all
  iterations, i.e. for no value of $\btheta_A$ does
  $\pi(\btheta_A\vert{\bf D}_A)\geq\pi(\btheta_A\vert{\bf D}_A^{(gen)})$. Thus, the null
  $H_0^{(A)}$ is rejected at a probability of 1. On the other hand,
  from the post-burnin part of the chain (beyond the 600,000-th
  iteration) we find that the minimum value of the posterior under the
  benchmark model ${\cal M}_{B}$ (shown in the black solid line) falls
  short of $\pi(\btheta_B\vert{\bf D}_B)$ at 87,650 number of
  iterations, out of the 200,000 samples of $\btheta_B$ generated in
  the post-burnin part of the chain run with ${\bf D}_B$.
 The null $H_0^{(B)}$ is then rejected at a probability of 1 - 87650/200,000 $\approx$ 0.4606.
}
\label{fig:sim_res}
\end{figure*}

\begin{table}
\noindent
\caption{Table displaying conditional probability of null $H_0^{(i)}$ (Statement~\ref{eqn:null_syn}) given synthetic data ${\bf D}_{i}$ that is simulated from true anisotropic state space density $f_i^{(True)}(\bX,\bV)$, where the density for $i=A$ is more anisotropic than for $i=B$. Column~2 shows the value $r_a$ of the anisotropy parameter that parametrises the deviation of $f_i^{(True)}(\bX,\bV)$ from an isotropic function of $\bX$ and $\bV$.
Column~3 shows the number $P_i$ of generated samples of $\btheta_i$ for which the posterior probability density given data ${\bf D}_i$, exceeds the minimum values of the posterior density under the null; column~4 gives the total number $Q_i$ of samples of $\btheta_i$ generated in the chain. The ratio of the entries in Column~3 to that in Column~4 is in Column~5--it is taken to approximate $\Pr(\btheta_i\in{\cal T}_{{\cal M}_i}({\bf D}_i))$ which in turn is equal to $\Pr(H_0^{(i)}\vert {\bf D}_i)$ (see Equation~\ref{eqn:T} and Equation~\ref{eqn:support}). 
Column~6 delineates the probability at which null $H_0^{(i)}$ can be rejected, given data ${\bf D}_i$.}
 \label{table:sim}
\begin{center}
{\footnotesize
\begin{tabular}{|c||c|c|c|c|c|}\hline
\hline
$i$ & $r_a$ (kpc) & $P_i$ & $Q_i$ & $\Pr(\btheta_i\in{\cal T}_{{\cal M}_i}({\bf D}_i))\approx P_i/Q_i$ & $H_0^{(i)}$ rejected at probability\\
\hline
A & 4 & 0 & 2$\times$10$^5$ & 0 & 1\\
B & 0.2 & 87,650 & 2$\times$10$^5$ & 0.5394 & 0.4606\\
\hline
\end{tabular}
}
\vspace{-0.1in}
\end{center}
\end{table}

\begin{figure*}[!bt]
\begin{center}
  {
\hspace{-1.5cm}
\includegraphics[height=7.5cm]{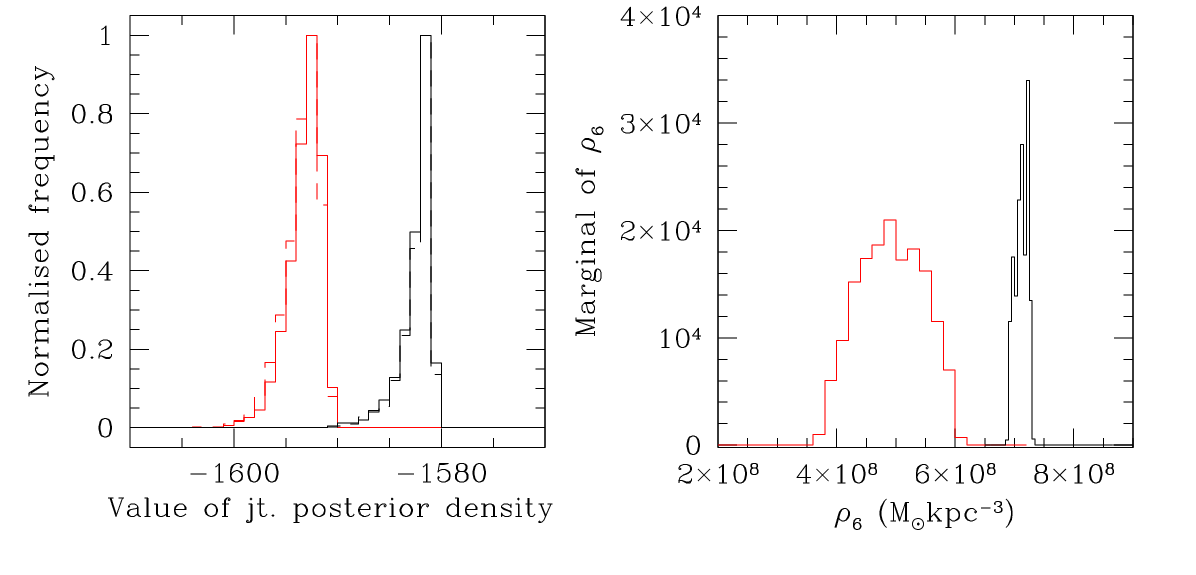} 
}
\end{center}
\caption{Left: figure showing histograms of the logarithm of the values of $\pi(\btheta_B\vert{\bf D}_B)$ generated in two distinct 30,000 iteration-long, post-burnin parts of the chain run with synthetic data ${\bf D}_B$ (histograms of values of the posterior in the two distinct parts, are shown in solid and broken lines coloured grey--or red in the e-version). Similar histograms of values of $\pi(\btheta_B\vert{\bf D}_B^{(gen)})$ generated in two distinct 30,000 iterations-long, post-burnin parts of the chain run with generated data ${\bf D}_B^{(gen)}$, are shown in solid and broken, black lines. Right: figure showing the marginal posterior probability of the parameter $\rho_6$, given synthetic data ${\bf D}_B$, plotted in grey, (or in red in the electronic version) and the marginal of $\rho_6$ given and data ${\bf D}_B^{(gen)}$ (in black), where ${\bf D}_B^{(gen)}$ is sampled from the isotropic state space $pdf$ that is itself learnt using ${\bf D}_B$. }
\label{fig:sim_2}
\end{figure*}

We corroborate convergence within the parts of the chains that we refer to as ``post-burnin'' in chains run with ${\bf D}_B$ and ${\bf D}_B^{(gen)}$ in Figure~\ref{fig:sim_2}, by overplotting histograms of values of joint posterior probability density--of $\btheta_B$ given the data--generated over two distinct but equally long parts of such post-burnin stage of the chains. Concurrence of these generated histograms offers confidence in the convergence achieved in the post-burnin stage of the chains presented in the left panel of Figure~\ref{fig:sim_res}. In Figure~\ref{fig:sim_2}, we also present the marginal densities of the parameter $\rho_6$, given synthetic data ${\bf D}_B$ (sampled from a chosen state $pdf$ that is mildly anisotropic) and generated data ${\bf D}_B^{(gen)}$ (sampled from the isotropic $pdf$ that is learnt using data ${\bf D}_B$).

\section{Testing for isotropic nature of state space of a real galaxy}
\label{sec:RESULTS}
\noindent
In Section~\ref{sec:application}, we introduced the main application that we
address in this work, namely that of learning the density function
of all gravitating mass in the real galaxy NGC 3379, using two
independent real data sets ${\bf D}_1$ observed by \ctn{bergond} and ${\bf
  D}_2$ observed by \ctn{douglas_07}. These are two distinct data sets that
bear information about 3--out of the 6--state space coordinates of two
different kinds of galactic particles, referred to as planetary
nebulae (PNe) and globular clusters (GCs). The data used in the work
include measurements of $X_1$, $X_2$ and $V_3$ of 164 PNe reported by
\ctn{douglas_07} and of 29 GCs by \ctn{bergond}. From the estimate of (the
discretised version $\brho$ of) the gravitational mass density function
of all types of matter in the galaxy, the mass density function of
luminous matter in the galaxy can be subtracted, leaving us the mass
density of the dark matter in the galaxy, which is a crucially
important input into cosmological models. See Section~\ref{sec:application}
for details.

As the learning of $\brho$ is possible only under the assumption
that the available data is sampled from an isotropic state space
density function, in this section, we discuss finding the probability
of the null that the state space of this example real galaxy is
isotropic, conditional on the measured data sets ${\bf D}_1$ and
${\bf D}_2$. Having estimated $\brho$ using ${\bf D}_1$ and then using
${\bf D}_2$, each time assuming that the galactic state space is
isotropic, we want to know in which case this assumption was more
invalid, given the data. In other words, we want to find the
comparative support for the null in these two data sets.

The physical implications of unequal supports for the assumption that
the state space of a given galaxy is isotropic, can be most
interesting--such would then imply that different sub-volumes of the
galactic state space are differently anisotropic. This in turn implies
that the state space of the galaxy is marked by at least two
non-interacting sub-volumes, the dynamical structures of which are
different, i.e. the distribution of the location $\bX$ and velocity
$\bV$ vectors of the galactic particles in which are different. The
non-linear dynamical implications of such difference is that the
motions of particles in these sub-volumes do not communicate. Physical
processes that cause such a split nature of the galactic state space
will then be sought, and importantly, it will then be acknowledged
that estimating the mass density of dark matter in a real galaxy using
the available measurements on $X_1,X_2,V_3$ of one set of galactic
particles--as is the usual practice in astrophysics--can be risky.

The null $H_0^{(i)}$, that data ${\bf D}_i$ is sampled from an
isotropic state space density function $\Psi_i(E(X,V))$ is defined in
Statement~\ref{eqn:nulldefn}; $i=1,2$. Our new test, as described in
Section~\ref{sec:implementation}, is implemented to estimate the
conditional probability $\Pr(H_0\vert{\bf D}_i)$ of the null
$H_0^{(i)}$ given the data ${\bf D}_i$. To compute this, we generate
data ${\bf D}_i^{(gen)}$ by rejection sampling from the discretised
state space $pdf$ that is itself learnt using measured data ${\bf
  D}_i$) under the benchmark model ${\cal M}_i$ (in which $H_0^{(i)}$
is true).

To compute $\Pr(H_0\vert{\bf D}_i)$, 3 chains: $i-RUN~I$,
$i-RUN~II$ and $i-RUN~III$, that are distinguished by the seeds or
initial guesses for the unknown parameters, are started with the
available galactic data ${\bf D}_i$, for $i=1,2$, with the aim of
learning the unknown model parameter vector
$\btheta_i=(\Psi_1^{(i)},\ldots,\Psi_{N_E}^{(i)},\rho_1^{(i)},\ldots,\rho_{N_x}^{(i)})^T$,
where the vector $\brho_i=(\rho_1^{(i)},\ldots,\rho_{N_x}^{(i)})^T$ is
the discretised version of the sought density function of
gravitational mass of all matter in the galaxy and
$\bPsi_i=(\Psi_1^{(i)},\ldots,\Psi_{N_E}^{(i)})^T$ is the discretised
version of the state space density $\Psi_i(E)$, as learnt using the
Bayesian scheme detailed in Section~{\bf S-1}
of the attached Supplementary Material, under the assumption that
${\bf D}_i$ is sampled from an isotropic state space density. The
chains are at least 800,000 iterations long, and the unknown model
parameter $\btheta_i$ is estimated using uniform priors on each scalar
unknown, $\Psi_j^{(i)}$ and $\rho_k^{(i)}$, are used,
$j=1,\ldots,N_E$, $k=1,\ldots,N_x$. From each chain, the identified
${\bPsi_i}^{(M,i)}$ at the identified $\brho_i^{(M,i)}$ is used to
generate a data set ${\bf D}_i^{(gen)}$ (see
Section~\ref{sec:implementation}). A chain is run with this generated
data set, in order to compute the minimal value of the posterior
when the null is true. For each of the three chains initiated with
different seeds and data ${\bf D}_i$, we identify the fractional
number of samples of $\btheta_i$ for which
$\pi(\btheta_i^{(min)}\vert {\bf D}_i^{(gen)}) \leq
\pi(\btheta_i\vert {\bf D}_i)$, for
each $i$=1,2.  The results for each chain are presented in
Table~\ref{table:radius2}.

\begin{table}
\caption{Table showing support in data ${\bf D}_i$ for null $H_0^{(i)}$, $i=1,2$, computed using 3 different chains $i-RUN~j$ for each $i$; $i=1,2$, $j=I,II,III$.}
 \label{table:radius2}
\begin{center}
{\footnotesize
\begin{tabular}{|c||c|c|}\hline
\hline
Chain name & Data set used & $\Pr(H_i\vert{\bf D}_i)$ \\
\hline
$1-RUN~I$ & ${\bf D}_1$ & 0.6202\\
$1-RUN~II$ & ${\bf D}_1$ & 0.5862\\
$1-RUN~III$ & ${\bf D}_1$ & 0.6269\\
$2-RUN~I$ & ${\bf D}_2$ & 0.9617\\
$2-RUN~II$ & ${\bf D}_2$ & 0.9650\\
$2-RUN~III$ & ${\bf D}_2$ & 0.9348\\
\hline
\end{tabular}
}
\vspace{-0.1in}
\end{center}
\end{table}

Traces of the log of the posterior probabiliy density of
$\btheta_i$ given real data ${\bf D}_i$ in the chains $i-RUN~I$, for
$i=1,2$ are shown in Figure~\ref{fig:hist_real}. The minimum value of the posterior density under the null $H_0^{(i)}$ is depicted in the solid line starting from the end of the burnin stage of the chain.

\begin{figure*}[!b]
\includegraphics[width=12cm]{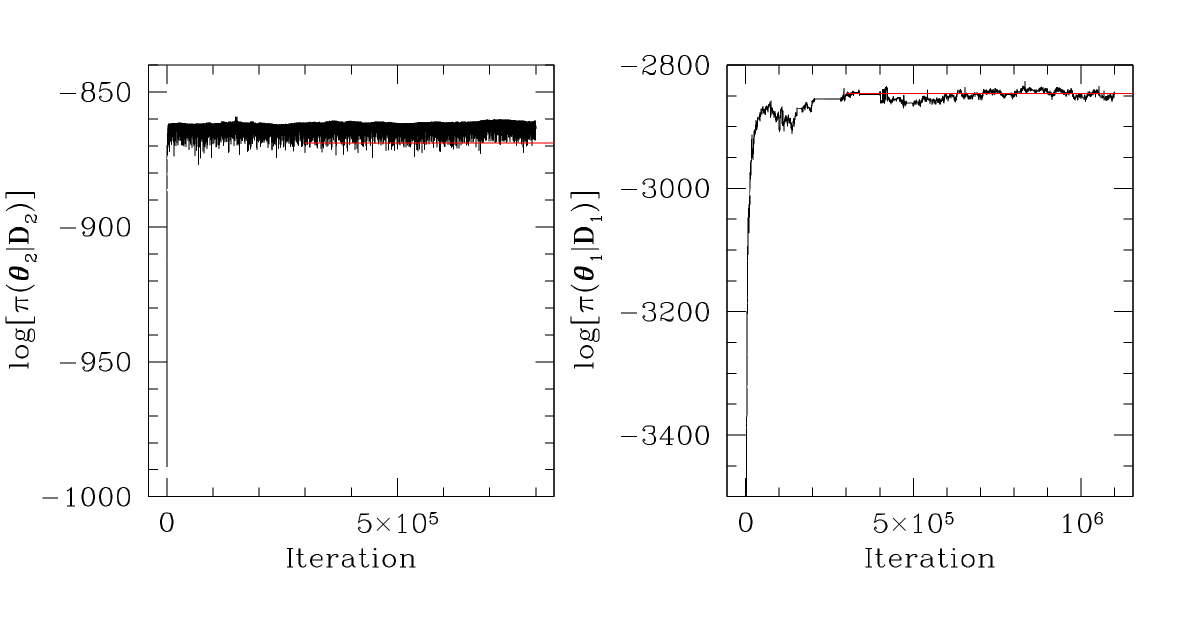}
\vspace{-1cm}
\caption{{\small Trace of logarithm of the posterior probability density of the model parameter vector ${\btheta}_1$ (right panel) and $\btheta_2$ (left) given the two sets of real data ${\bf D}_1$ (size 164) and ${\bf D}_2$ (size 29) respectively, in chains $1-RUN~I$ and $2-RUN~I$. The minimal value of the posterior under the benchmark model (when the null is true given the corresponding generated data set), from the post-burnin stage of that chain (iteration 300,000 onwards), is shown in the solid grey (or red in the e-version) line.}}
\label{fig:hist_real}
\end{figure*}

\begin{figure*}[!t]
\includegraphics[width=14cm]{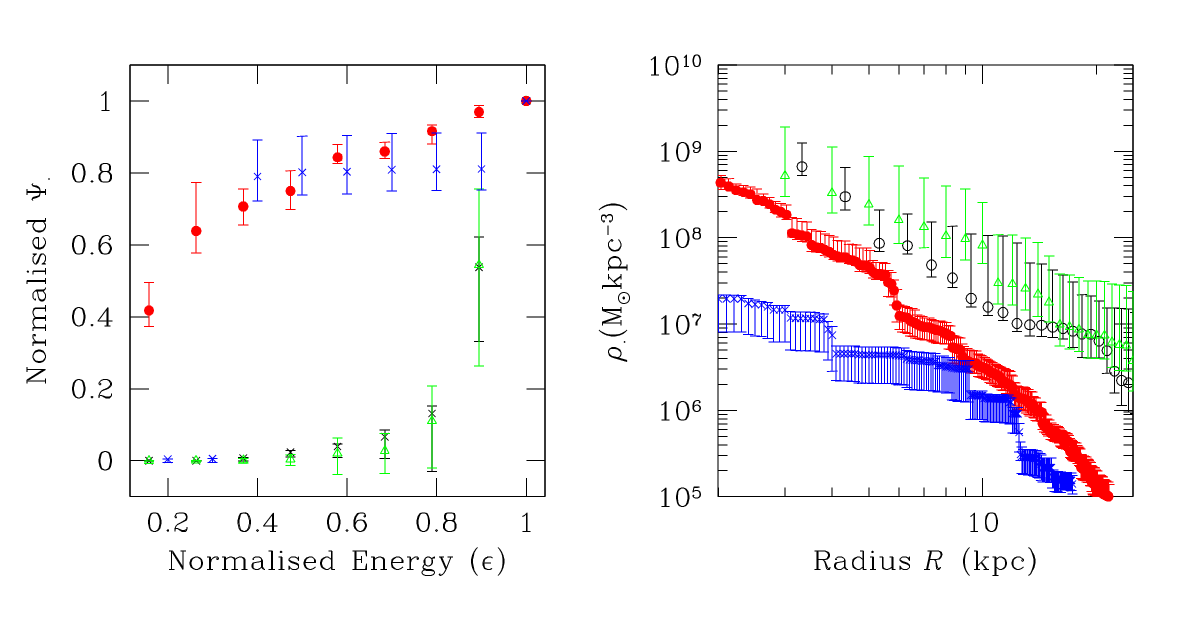}
\vspace{-1cm}
\caption{\label{fig:fig2} {\small Right panel: logarithm of
    gravitational mass density vector $\brho_{2}$ (in black, with
    modal values shown in open circles) learnt from chain $2-RUN~I$ that is
    run using data ${\bf D}_2$, and $\brho_{1}$ from chain $1-RUN~I$
    that is run using ${\bf D}_1$ (modal values shown in filled circles; in
    red in the e-version). These gravitational mass density results
    were obtained under the assumption of an isotropic state space,
    the support for which in the two data sets is indicated in
    Table~\ref{table:radius2}. Overlaid on these are the identified vectors
    $\brho_1^{(min)}$ (modal values in crosses; in blue in the
    e-version) and $\brho_2^{(min)}$ (modal values in triangles; in
    green in the e-version), which are respectively, the
    posterior-minimising, null-abiding, gravitational mass density
    vectors identified in chains run with the generated data ${\bf
      D}_1^{(gen)}$ and ${\bf D}_2^{(gen)}$. The concurrence of
    $\brho_{2}$ and $\brho_2^{(min)}$ is noted, along with the lack
    of consistency between $\brho_{1}$ and $\brho_1^{(min)}$. The
    error bars represent the 95$\%$ HPD credible regions on the
    estimated $\rho_\cdot$ parameter. In the left panel, the state
    space density vectors ${\bPsi}_{1}$ (modal values in filled red
    circles) and ${\bPsi}_{2}$ (modal values in open black circles),
    learnt from the chains $1-RUN~I$ and $2-RUN~I$, are shown,
    compared respectively to ${\bPsi}_1^{(min)}$ (modal values in
    blue crosses) and ${\bPsi}_2^{(min)}$ (in green
    triangles). Again, the overlap of ${\bPsi}_{(2)}$ and
    ${\bPsi}_2{(min)}$ is noted, as is the discord between
    ${\bPsi}_{1}$ and ${\bPsi}_1^{(min)}$, especially at high
    and low energies. The ${\bPsi}$ vectors are normalised to unity at
    $\epsilon=1$ where $\epsilon$ is the value of the normalised
    energy.}}
\end{figure*}

Basically, support in real data ${\bf D}_1$ for the assumption of an
isotropic state space, is distinct from that in ${\bf D}_2$. This
implies that the $f_1({\bx, \bv}) \neq f_2({\bx, \bv})$, where the
true state space $pdf$ that ${\bf D}_1$ is sampled from is $f_1({\bx,
  \bv})$ and ${\bf D}_2\sim f_2({\bx, \bv})$. However, both data sets
carry information on the state space coordinates
$(X_1,X_2,X_3,V_1,V_2,V_3)^T$ in the same galactic state space,
i.e. both data sets are sampled from $pdf$s that describe the state
space structure of all or some volume inside the same galactic state
space ${\cal W}$. Thus, $f_1({\bx, \bv}) \neq f_2({\bx,
  \bv})\Longrightarrow {\cal W}_1\neq {\cal W}_2$ where $f_1({\bx,
  \bv})$ is the $pdf$ of the state space vector that lives in volume
${\cal W}_1\subset{\cal W}$ and $f_2({\bx, \bv})$ is the density of
the state space vector in volume ${\cal W}_2\subset{\cal W}$. In terms
of the state space structure of this real galaxy NGC~3379, we can then
conclude that the state space of the system is marked by at least two
distinct volumes, motions in which do not communicate with each other,
leading to distinct particle distributions being set up in these two
volumes, which in turn manifests in distinct $pdf$s for these
subspaces (${\cal W}_1$ and ${\cal W}_2$) of the galactic state space
${\cal W}$. Data ${\bf D}_1$ and ${\bf D}_2$ are respectively drawn
from such distinct $pdf$s.

Comparing the computed $\Pr(H_0^{(1)}\vert{\bf D}_1)$ and
$\Pr(H_0^{(2)}\vert{\bf D}_2)$, we can see that the assumption of
isotropy is more likely to be invalid for the state space density from
which the data ${\bf D}_1$ are sampled than from which the data ${\bf
  D}_2$ are drawn.  Even beyond comparative terms, our results
indicate that $\Pr(H_0^{(2)}\vert{\bf D}_2)\approx 1$, i.e. we reject the isotropy of the state space density that the observed data ${\bf
  D}_2$ in this galaxy live in at nearly 0 probability.




\section{Discussions}
\label{sec:discussions}
\noindent
In the above test, a high support in ${\bf D}_2$ towards an isotropic
state space $pdf$, along with a moderate support in ${\bf D}_1$ for
the same assumption, indicate that the two samples are drawn from two
distinct state space densities.

Any apriori expectation that the implementation of the PNe and GC data
sets will lead to concurring gravitational mass density estimates is
foreshadowed by the assumption that both data sets are sampled from
the same - namely, the galactic - state space density $f(\bX,\bV)$.
Such an expectation can be understood to emanate from the argument
that since both samples live in the galactic phase space ${\cal W}$,
they are expected to be sampled from the same galactic state space
density, at the galactic gravitational potential. However, such does
not necessarily follow if--for example--the galactic state space
density $f({\bX,\bV})$ is a non-analytic function with $p_{max}$
branches:
\begin{equation}
f({\bX,\bV})=f_p({\bX,\bV}), \quad \forall{(X_1,X_2,X_3,V_1,V_2,V_3)^T }\in{\cal W}_p\subseteq{\cal W},\quad p=1,\ldots,p_{max}.
\end{equation}
Then, if the data ${\bf
  D}_2$ are sampled from the density $f_2(\cdot)$ and data
${\bf D}_1\sim f_1(\cdot)$, it follows that
${\bf D}_1$ and ${\bf D}_2$ are sampled from unequal
state space densities.
Qualitatively we understand that if the galactic state space ${\cal
  W}$ is split into isolated volumes, such that the motions in these
volumes do not mix and are therefore distinctly distributed in
general, the state space densities of these volumes would be
unequal. This is synonymous to saying that ${\cal W}$ is marked by at
least two distinct basins of attraction and the two observed samples
reside in such distinct basins.

One standard non-linear dynamical cause for the splitting of ${\cal
  W}$ include the development of basins of attraction, leading to
attractors, generated in a multistable galactic gravitational
potential. Basins of attraction could also be triggered around chaotic
attractors, which in turn could be due to resonance interaction with
external perturbers or due to merging events in the evolutionary
history of the galaxy.  Galactic state spaces can be split given that
a galaxy is expectedly a complex system, built of multiple components
with independent evolutionary histories and distinct dynamical
timescales. As an example, at least in the neighbourhood of the Sun,
the state space structure of the Milky Way is highly multi-modal and
the ensuing dynamics is highly non-linear, marked by significant
chaoticity.
}

\begin{center}
{\bf{Supplementary material}}
\end{center}
{Details of the Bayesian learning of the gravitational mass density
  and state space $pdf$ of the galaxy are provided in Section
  {\bf{S-1}} of the attached supplementary material.  Section
  {\bf{S-2}} discusses details of the Fully Bayesian Significance
  Test.}

\section*{Acknowledgments}
I gratefully acknowledge the comments of the reviewers that helped
improve the paper.

\renewcommand\baselinestretch{1.3}


\end{document}